\documentclass[useAMS,usenatbib]{mn2e}
\usepackage{amsmath}
\usepackage{amssymb}
\usepackage{amsfonts}
\usepackage{xcolor}
\usepackage{graphicx}
\usepackage{epstopdf}
\usepackage{enumerate}
\usepackage{subfigure}
\usepackage[hyperindex,breaklinks=true, colorlinks, citecolor=blue]{hyperref} 

\def\deg{\ifmmode^\circ\else$^\circ$\fi}

\def\ba{\begin{eqnarray}}
\def\ea{\end{eqnarray}}
\def\be{\begin{equation}}
\def\ee{\end{equation}}

\makeatletter
\def\ref@jnl#1{{\rmfamily#1}}%
\newcommand\aj{\ref@jnl{AJ}}%
\newcommand\araa{\ref@jnl{ARA\&A}}%
\newcommand\apj{\ref@jnl{ApJ}}%
\newcommand\apjl{\ref@jnl{ApJ}}%
\newcommand\apjs{\ref@jnl{ApJS}}%
\newcommand\apss{\ref@jnl{Ap\&SS}}%
\newcommand\aap{\ref@jnl{A\&A}}%
\newcommand\aapr{\ref@jnl{A\&A~Rev.}}%
\newcommand\aaps{\ref@jnl{A\&AS}}%
\newcommand\baas{\ref@jnl{BAAS}}%
\newcommand\memras{\ref@jnl{MmRAS}}%
\newcommand\mnras{\ref@jnl{MNRAS}}%
\newcommand\pra{\ref@jnl{Phys.~Rev.~A}}%
\newcommand\prb{\ref@jnl{Phys.~Rev.~B}}%
\newcommand\prc{\ref@jnl{Phys.~Rev.~C}}%
\newcommand\prd{\ref@jnl{Phys.~Rev.~D}}%
\newcommand\pre{\ref@jnl{Phys.~Rev.~E}}%
\newcommand\prl{\ref@jnl{Phys.~Rev.~Lett.}}%
\newcommand\pasp{\ref@jnl{PASP}}%
\newcommand\pasj{\ref@jnl{PASJ}}%
\newcommand\ssr{\ref@jnl{Space~Sci.~Rev.}}%
\newcommand\nat{\ref@jnl{Nature}}%
\newcommand\iaucirc{\ref@jnl{IAU~Circ.}}%
\newcommand\aplett{\ref@jnl{Astrophys.~Lett.}}%
\newcommand\apspr{\ref@jnl{Astrophys.~Space~Phys.~Res.}}%
\newcommand\nphysa{\ref@jnl{Nucl.~Phys.~A}}%
\newcommand\physrep{\ref@jnl{Phys.~Rep.}}%
\newcommand\planss{\ref@jnl{Planet.~Space~Sci.}}%
\newcommand\procspie{\ref@jnl{Proc.~SPIE}}%

\makeatletter
\newcommand\footnoteref[1]{\protected@xdef\@thefnmark{\ref{#1}}\@footnotemark}
\makeatother

\title[Extracting HI signal with {\tt GNILC}]{Extracting HI cosmological signal with Generalized Needlet Internal Linear Combination}
\author[L.\,C.\,Olivari, M.\,Remazeilles, and C.\,Dickinson]{L.\,C.\,Olivari\thanks{\url{E-mail: lucas.olivari@postgrad.manchester.ac.uk}}$^1$, M.\,Remazeilles\thanks{\url{E-mail: mathieu.remazeilles@manchester.ac.uk}}$^1$, C.\,Dickinson\thanks{\url{E-mail: clive.dickinson@manchester.ac.uk}}$^1$ \\
$^1$Jodrell Bank Centre for Astrophysics, Alan Turing Building, School of Physics \& Astronomy, The University of Manchester, \\
Oxford Road, Manchester, M13 9PL, U.K. \\
}
\begin{document}


\setlength{\topmargin}{-15mm}

\pagerange{\pageref{firstpage}--\pageref{lastpage}} \pubyear{2002}

\maketitle

\label{firstpage}

\begin{abstract}
HI intensity mapping is a new observational technique to map fluctuations in the large-scale structure of matter using the 21 cm emission line of atomic hydrogen (HI). Sensitive HI intensity mapping experiments have the potential to detect Baryon Acoustic Oscillations (BAO) at low redshifts ($z \lesssim 1$) in order to constrain the properties of dark energy. Observations of the HI signal will be contaminated by instrumental noise and, more significantly, by astrophysical foregrounds, such as Galactic synchrotron emission, which is at least four orders of magnitude brighter than the HI signal. Foreground cleaning is recognised as one of the key challenges for future radio astronomy surveys. We study the ability of the Generalized Needlet Internal Linear Combination ({\tt GNILC}) method to subtract radio foregrounds and to recover the cosmological HI signal for a general HI intensity mapping experiment. The {\tt GNILC} method is a new technique that uses both frequency and spatial information to separate the components of the observed data. Our results show that the method is robust to the complexity of the foregrounds. For simulated radio observations including HI emission, Galactic synchrotron, Galactic free-free, radio sources and $0.05$\,mK thermal noise, we find that the {\tt GNILC} method can reconstruct the HI power spectrum for multipoles $30 < \ell < 150$ with 6$\%$ accuracy on 50$\%$ of the sky for a redshift $z\sim 0.25$.

\end{abstract}

\begin{keywords}
methods: data analysis -- radio continuum: general, galaxies -- radio lines: ISM -- cosmology: observations -- large-scale structure of Universe

\end{keywords}

\section{Introduction}
\label{sec:intro}

There are several observational methods that can be used to constraint the properties of the dark energy, one of the most powerful of them is the Baryon Acoustic Oscillations (BAO) \citep{Weinberg2013}. Until now, all of the BAO observations were made using redshift surveys performed in the optical and near infrared wavebands or using the $3$-dimensional structure in the Ly$\alpha$ forest absorption towards a dense grid of high-redshift quasars \citep{Aubourg2014}. Present sample sizes for the detection of BAO are typically $10^5$ - $10^6$, but there are a number of projects planned, such as Euclid \citep{Laureijs2011} and LSST \citep{LSST2012}, to increase this to $10^7$ - $10^9$ using optical and near-infrared observations. It is possible, on the other hand, to perform redshift surveys in the radio waveband using 21 cm radiation from neutral hydrogen (HI) to select galaxies \citep{Chang2008}. The use of a different waveband could improve the confidence in the results from the optical and near-infrared surveys.

One way to do radio redshift surveys is through a technique called HI intensity mapping \citep*{Madau1997ApJ, Battye2004, Peterson2006, Loeb2008}. Intensity mapping is the study of the large-scale fluctuations in the intensity of a given spectral line emitted by a number of unresolved objects. It has been suggested that the full intensity field could be used to measure the power spectrum as a function of redshift if the continuum emission, such as the one from our Galaxy, can be accurately subtracted and any systematic instrumental noise can be calibrated to sufficient accuracy \citep{Peterson2009}. The main advantage of HI intensity mapping, compared to the optical surveys of galaxies, is that a large sky volume is achieved within a relatively short observing time.

Using the HI intensity mapping method, the Green Bank Telescope (GBT) has made the first detection of the HI signal at $z \approx 0.8$. To have enough significance in the detection of the faint HI signal, \citet{Masui2013} cross-correlated the 21 cm data obtained with the GBT with the optical data obtained by the WiggleZ Dark Energy Survey. This detection showed that the HI intensity mapping is a very promising tool to study the large scale structure of the Universe. 

It is worth stressing that HI intensity mapping is a new observational technique that is still being developed. Ongoing or planned observational efforts, such as CHIME  \citep{Bandura2014}, BINGO \citep{Battye2013}, TIANLAI \citep{Chen2012}, FAST \citep{Smoot2014} and SKA \citep{Santos2015}, are essential for making this technique a competitive probe of the physical properties of the Universe.

There are two main techniques for performing an HI intensity mapping experiment: single-dish and interferometer arrays. Both kinds of experiment will have to deal with potential systematics, such as gain variations and correlated noise in frequency. Single dish experiments will require stable receiver systems and good calibration.  Interferometers, on the other hand, are known to deal more naturally with systematics than single-dishes \citep{Dickinson2012, White1999}. It is worth noting, however, that existing interferometers are limited by the small number of their smallest baselines and therefore do not provide the required surface brightness sensitivity \citep{Bull2014}.

As already mentioned, the success of any HI intensity mapping experiment strongly depends on our ability to subtract the astrophysical contaminations that will be present in the observed HI signal. At $\sim 1$ GHz, the most relevant foregrounds are the Galactic emission, mostly synchrotron radiation, and the background emission of extragalactic point sources. These emissions are at least four orders of magnitude larger, $T_b \sim 10$ K, than the HI signal, $T_b \sim 1$ mK. The high spectral resolution offered by any HI intensity mapping experiment allows us to use the frequency information of the observed data. As the foregrounds spectra are expected to be smooth, we can approximate them by a power-law in the frequency range of interest. This property can then be used to separate the HI signal from any other signal correlated in frequency \citep{Ansari2012, Liu2012}.

There are other component separation techniques available in the literature for HI intensity mapping experiments, both in the reionisation epoch and for low redshifts. Some of them, such as Karhunen-Loeve Decomposition \citep{Shaw2014}, are parametric methods. Parametric methods use a model to describe some physical properties of the foregrounds. Others, such as Principal Component Analysis ({\tt PCA}) \citep{Masui2013, Switzer2013, Alonso2015, Bigot-Sazy2015}, Independent Component Analysis ({\tt ICA}) \citep{Alonso2015}, {\tt FASTICA} \citep{Wolz2014}, inverse variance \citep{Liu2011}, and quadratic estimators \citep{Switzer2015}, use only the observed data to recover the HI signal and therefore do not assume a specific parametric model for the foregrounds.

In this work, we study the application of the Generalized Needlet Internal Linear Combination ({\tt GNILC}) \citep{Remazeilles2011b} as a non-parametric component separation technique for HI intensity mapping experiments. In general, the {\tt GNILC} method can extract the emission of a multidimensional component (spatially correlated components) from the observed data. Originally, \citet{Remazeilles2011b} used the {\tt GNILC} method to obtain the total emission of the Galactic foregrounds from CMB data. Here we apply the {\tt GNILC} method to a single-dish HI intensity mapping in low redshifts.

We organize this paper as follows. In Section~\ref{sec:method}, we present the formalism of the {\tt GNILC} method and show how it can be used in an HI intensity mapping experiment. In Section~\ref{sec:simu}, we describe the models that we use to simulate the different components of the observed sky. In Section~\ref{sec:results}, we describe and discuss the results that we have obtained. Finally, in Section~\ref{sec:conclusions}, we make our final remarks about the present work.

\section{GNILC method}
\label{sec:method}

The Internal Linear Combination (ILC) method was first used, in the CMB context, by the WMAP team in \citet{Bennett2003}. The ILC method applies to the observed data a weight matrix $W$ that offers unit response to the CMB emission while it minimizes the total variance of the foreground plus noise signal.

\citet{Bennett2003} used the ILC method in pixel space. Later, the ILC method was used in harmonic space by \citet{Tegmark2003} and in wavelet space by \citet{Delabrouille2009}. Finally, \citet*{Remazeilles2011b} developed an extension of the ILC method in wavelet space that makes use of a prior for the power spectrum of a specific component of the observed sky. This extension of the ILC method is called Generalized Needlet Internal Linear Combination ({\tt GNILC}). Here, we adapt the {\tt GNILC} method to HI intensity mapping experiments.

We model the sky observation, $x_i(p)$, at frequency $i$ and pixel (or direction of the sky) $p$, as

\begin{equation}
\label{data}
x_i(p) = s_i(p) + n_i(p),
\end{equation} where $s_i(p)$ is the HI signal and $n_i(p)$ is the astrophysical foregrounds plus instrumental noise contribution to the observed data. Eq. $\eqref{data}$ can be rewritten in the $n_{\mathrm{ch}} \times 1$ vector form, where $n_{\mathrm{ch}}$ is the number of frequency channels,

\begin{equation}
\textbf{\textit{x}}(p) = \textbf{\textit{s}}(p) + \textbf{\textit{n}}(p).
\end{equation} Here $\textbf{\textit{x}}$ are the $n_{\mathrm{ch}}$ observation maps, each being a mixture of the HI signal $\textbf{\textit{s}}$ and the foregrounds plus noise component $\textbf{\textit{n}}$.

The $n_{\mathrm{ch}} \times n_{\mathrm{ch}}$ covariance matrix of the sky observations, $\mathbfss{R}(p) = \langle \textbf{\textit{x}}(p)\textbf{\textit{x}}^T(p) \rangle$, at pixel $p$, is

\begin{equation}
\label{ocov}
\mathbfss{R}(p) = \mathbfss{R}_{\mathrm{HI}}(p) + \mathbfss{R}_n(p),
\end{equation} where $\mathbfss{R}_{\mathrm{HI}}(p) = \langle \textbf{\textit{s}}(p)\textbf{\textit{s}}^T(p) \rangle$ is the covariance matrix of the HI emission and $\mathbfss{R}_n(p) = \langle \textbf{\textit{n}}(p)\textbf{\textit{n}}^T(p) \rangle$ is the covariance matrix of the foregrounds plus noise.

The foregrounds plus noise signal may significantly vary with the observed directions in the sky. The relative power ratio between foregrounds, noise, and HI signals also changes with the angular scale considered. Therefore, to describe theoretically the foreground emissions of the sky, we need a large number of degrees of freedom; this number of degrees of freedom would be infinite in the case of a infinitely narrow beam. However, in practice, to describe the observed data, we are limited by the number of frequency channels $n_{\mathrm{ch}}$ of our experiment. Moreover, the foreground components of emission are correlated over frequencies, so that in practice the foregrounds plus noise signal can be represented as a linear combination of a finite number $m$ of independent (unphysical) templates. The HI signal $\textbf{\textit{s}}$ is also partially correlated over adjacent frequencies (redshift bins), therefore it can also be described by a finite number of degrees of freedom, which in practice would be given by $n_{\mathrm{ch}} - m$, where $m$ is the dimension of the foreground subspace. In this way, the HI signal can be represented as the superposition of $n_{\mathrm{ch}} - m$ independent (unphysical) templates $\textbf{\textit{t}}$,

\begin{equation}
\label{st}
\textbf{\textit{s}} = \mathbfss{S} \textbf{\textit{t}},
\end{equation} where $\mathbfss{S}$ is an $n_{\mathrm{ch}} \times (n_{\mathrm{ch}} - m)$ mixing matrix giving the contribution from the templates to the HI emission in each frequency channel. We note that the templates $\textbf{\textit{t}}$ are not physical templates; instead they allow us to explore the $(n_{\mathrm{ch}} - m) \times (n_{\mathrm{ch}} - m)$ submatrix of the observation covariance matrix that is dominated by the HI signal. Therefore, using Eq. $\eqref{st}$, we see that the covariance matrix of the HI signal is given by an $n_{\mathrm{ch}} \times n_{\mathrm{ch}}$ matrix with rank equal to $n_{\mathrm{ch}} - m$,

\begin{equation}
\mathbfss{R}_{\mathrm{HI}} = \mathbfss{S} \mathbfss{R}_t \mathbfss{S}^T,
\end{equation} where $\mathbfss{R}_t = \langle \textbf{\textit{t}} \textbf{\textit{t}}^T \rangle$ is a full-rank $(n_{\mathrm{ch}} - m) \times (n_{\mathrm{ch}} - m)$ matrix.

Now, we consider the estimation of $\textbf{\textit{s}}$ by a linear operation

\begin{equation}
\widehat{\textbf{\textit{s}}} = \mathbfss{W} \textbf{\textit{x}}, 
\end{equation} where the $n_{\mathrm{ch}} \times n_{\mathrm{ch}}$ ILC weight matrix $\mathbfss{W}$ offers unit response to the HI emission while minimizing the total variance of the vector estimate $\widehat{\textbf{\textit{s}}}$. This means that the matrix $\mathbfss{W}$ minimizes $E(\vert \mathbfss{W} \textbf{\textit{x}} \vert^2)$ under the constraint $\mathbfss{W} \mathbfss{S} = \mathbfss{S}$. The weight matrix $\mathbfss{W}$ thus solves the following minimization problem

\begin{equation}
\mathrm{min} \left[ \mathrm{Tr} ( \mathbfss{W} \mathbfss{R} \mathbfss{W}^T ) \right] \;\;\; \mathrm{with} \;\;\; \mathbfss{W} \mathbfss{S} = \mathbfss{S},
\end{equation} where $\mathbfss{R}$ is the covariance matrix of observations $\textbf{\textit{x}}$. This minimization problem, which is a multidimensional ILC problem, can be solved through the use of a Lagrange multiplier, which gives

\begin{equation}
\label{multi}
\mathbfss{W} = \mathbfss{S} (\mathbfss{S}^T \mathbfss{R}^{-1} \mathbfss{S} )^{-1} \mathbfss{S}^T \mathbfss{R}^{-1}.
\end{equation} It is important to notice that expression $\eqref{multi}$ for $\mathbfss{W}$ is invariant if $\mathbfss{S}$ is changed into $\mathbfss{S} \mathbfss{T}$ for any invertible matrix $\mathbfss{T}$. Hence, to implement the ILC filter Eq. $\eqref{multi}$, we only need to know the mixing matrix $\mathbfss{S}$ up to a multiplication by an invertible factor \citep{Remazeilles2011b}. In other words, we do not need to know the exact HI mixing matrix to perform component separation, but only the column space of $\mathbfss{S}$, i.e. the dimension $n_{\mathrm{ch}} - m$ of the HI subspace. Since the mixing matrix of the HI emission is unknown, this is a major advantage of the ILC filter compared to Wiener filters.

As we have mentioned, the mixing matrix $\mathbfss{S}$ is not known beforehand. Therefore, to be able to use the ILC filter given by Eq. $\eqref{multi}$, we need a method to estimate it, or more precisely to estimate its column space. To do this, we will perform a constrained version of the Principal Component Analysis ({\tt PCA}) to the observation covariance matrix. Our constrained {\tt PCA} algorithm, differently from the standard {\tt PCA} algorithm \citep{Jolliffe2002}, will be driven by the local signal to noise ratio (ratio between the HI power and the total observation power) by making use of a prior on the HI power spectrum. No assumption is made on the foregrounds. The signal to noise ratio will be computed locally both in pixel space and in harmonic space by performing a wavelet decomposition of the data. This constrained {\tt PCA} step in the {\tt GNILC} algorithm will allow us to estimate, on different directions in the sky and on different angular scales, the local number of foregrounds plus noise degrees of freedom (principal components) and the local number of HI degrees of freedom (complementary components). To determine the dimension of the HI (resp. foregrounds plus noise) subspace, we will also use a statistical information criterion to discriminate between the dominant eigenvalues that are due to the principal components and the eigenvalues that are due to HI degrees of freedom. Note that the prior on the HI power spectrum is only used for determining the dimension of the HI subspace at the constrained {\tt PCA} step, not at the ILC filtering step.

\subsection{Estimation of the observation covariance matrix and the HI covariance matrix}

The covariance matrix of the observations for each pair of frequency channels $a$ and $b$ can be computed by the following expression,

\begin{equation}
\label{covobs}
\mathbfss{R}_{ab}(p) = \sum_{p' \in \mathcal{D}(p)}  x_a(p') x_b(p'),
\end{equation} where $\mathcal{D}(p)$ is a domain of pixels centred around the pixel $p$. The choice of the domain $\mathcal{D}(p)$ is done in such a way that we are able to avoid artificial correlations between the HI signal and the foregrounds plus noise signal (see Appendix \ref{app:needlets}).

In order to estimate the dimension (the column space) of the HI mixing matrix $\mathbfss{S}$ used in the multidimensional ILC filter, Eq. $\eqref{multi}$, we adopt a prior on the HI power spectrum. By describing the statistics of the HI emission with its covariance matrix, we are assuming that the HI emission is described by a Gaussian field. This is a reasonable assumption because any departure from Gaussianity of the HI emission is negligible compared to the large non-Gaussianity of the Galactic foreground emission. 

Using a theoretical template (prior) of the HI emission angular power spectra for each pair of frequency channels, $C_{\ell}^ {a b}$, we simulate HI maps for the different frequency channels. From the simulated HI maps we compute, as in Eq. $\eqref{covobs}$, the coefficients in real space of the HI covariance matrix $\widehat{\mathbfss{R}}_{\mathrm{HI}\,ab}$ for each pair of frequencies. 

In this work, for the theoretical prior, we use the model implemented in the software CORA \citep{Shaw2014} for the HI power spectrum; this model, which considers only the fluctuations on the HI energy density, is discussed in section \ref{subsec:hi}. As we show in section \ref{subsec:prior}, the {\tt GNILC} method is not critically sensitive to the exact shape of the HI power spectrum but depends on the relative strength of the HI signal compared to the observed data; the prior HI power spectrum only serves to determine the effective dimension of the HI subspace (section \ref{subsec:cpca}).

The computation of the observation covariance matrix and the HI covariance matrix is made independently on the different wavelet scales (ranges of multipoles) considered (see Appendix \ref{app:needlets}). The use of wavelets allows us to determine the relative strength of the HI power with respect to the foregrounds power both on localized areas of the sky and on defined ranges of angular scales.

\subsection{Determination of the HI signal subspace with constrained PCA}
\label{subsec:cpca}

To determine the HI signal subspace on each wavelet domain considered (i.e. a given pixel domain and a given range of angular scales), we perform, using the estimate of the HI covariance matrix $\widehat{\mathbfss{R}}_{\mathrm{HI}}$, a constrained {\tt PCA} that is driven by the local signal to noise ratio.

First, we apply the following transformation to the frequency observations,

\begin{equation}
\label{white}
\textbf{\textit{x}} \rightarrow \widehat{\mathbfss{R}}_{\mathrm{HI}}^{-1/2} \textbf{\textit{x}},
\end{equation} such that the covariance matrix of the transformed observations becomes

\begin{equation}
\label{transcov}
\widehat{\mathbfss{R}}_{\mathrm{HI}}^{-1/2} \mathbfss{R} \widehat{\mathbfss{R}}_{\mathrm{HI}}^{-1/2}.
\end{equation} Using Eq. $\eqref{ocov}$, the covariance matrix of the transformed observations can be decomposed as

\begin{equation}
\label{tcov}
\widehat{\mathbfss{R}}_{\mathrm{HI}}^{-1/2} \mathbfss{R} \widehat{\mathbfss{R}}_{\mathrm{HI}}^{-1/2} = \widehat{\mathbfss{R}}_{\mathrm{HI}}^{-1/2}  \mathbfss{R}_n \widehat{\mathbfss{R}}_{\mathrm{HI}}^{-1/2} + \widehat{\mathbfss{R}}_{\mathrm{HI}}^{-1/2} \mathbfss{R}_{\mathrm{HI}} \widehat{\mathbfss{R}}_{\mathrm{HI}}^{-1/2}. 
\end{equation} Assuming that the prior HI covariance matrix $\widehat{\mathbfss{R}}_{\mathrm{HI}}$ is close to the real HI covariance matrix $\mathbfss{R}_{\mathrm{HI}}$, Eq. $\eqref{tcov}$ becomes

\begin{equation}
\label{power}
\widehat{\mathbfss{R}}_{\mathrm{HI}}^{-1/2} \mathbfss{R} \widehat{\mathbfss{R}}_{\mathrm{HI}}^{-1/2} = \widehat{\mathbfss{R}}_{\mathrm{HI}}^{-1/2} \mathbfss{R}_n \widehat{\mathbfss{R}}_{\mathrm{HI}}^{-1/2} + \tilde{\mathbfss{I}}, 
\end{equation} such that the power of the transformed HI signal is given by the matrix $\tilde{\mathbfss{I}}$, which is close to the identity matrix $\mathbfss{I}$, since $\widehat{\mathbfss{R}}_{\mathrm{HI}}^{-1/2} \mathbfss{R}_{\mathrm{HI}} \widehat{\mathbfss{R}}_{\mathrm{HI}}^{-1/2} \simeq \mathbfss{I}$.

By diagonalizing the transformed observation covariance matrix (Eq. $\eqref{power}$), we can separate the degrees of freedom of the foregrounds plus noise from the degrees of freedom of the HI signal. Eq. $\eqref{power}$ shows us that the HI signal has approximately unitary power. Thus, the HI subspace will be defined by the subset of eigenvectors corresponding to the eigenvalues of the transformed observation covariance matrix that are approximately equal to 1. Taking this in consideration, we obtain the following eigenstructure for the transformed observation covariance matrix,

\begin{align}
\label{diag}
\widehat{\mathbfss{R}}_{\mathrm{HI}}^{-1/2} &\mathbfss{R} \widehat{\mathbfss{R}}_{\mathrm{HI}}^{-1/2} = \nonumber \\ &[\mathbfss{U}_N \, \mathbfss{U}_S] \, \times  \, \begin{bmatrix} \lambda_1 + 1 & & & \\ & \cdots & &  \\ & & \lambda_m +1 & \\ & & & \mathbfss{I}  \\
\end{bmatrix} \, \times  \, \begin{bmatrix} \mathbfss{U}_N^T  \\ \mathbfss{U}_S^T \end{bmatrix},
\end{align} where for simplicity we take $\tilde{\mathbfss{I}} = \mathbfss{I}$. The eigenvalues of the covariance matrix $\widehat{\mathbfss{R}}_{\mathrm{HI}}^{-1/2} \mathbfss{R} \widehat{\mathbfss{R}}_{\mathrm{HI}}^{-1/2}$ that are approximately equal to $1$, therefore, contain the power of the HI signal, with the corresponding eigenvectors spanning the HI subspace. In this representation, the subset of eigenvectors that form the $n_{\mathrm{ch}} \times (n_{\mathrm{ch}} - m)$ matrix $\mathbfss{U}_S$ defines the independent templates that contribute to the HI signal. Conversely, the number $m$ of eigenvalues significantly larger than $1$ corresponds to the dimension of the foregrounds plus noise subspace, which is spanned by the set of eigenvectors collected in the matrix $\mathbfss{U}_N$.

We can write Eq. $\eqref{diag}$ in the following compact form

\begin{equation}
\label{eigenstructure}
\widehat{\mathbfss{R}}_{\mathrm{HI}}^{-1/2} \mathbfss{R} \widehat{\mathbfss{R}}_{\mathrm{HI}}^{-1/2} = \mathbfss{U}_N \mathbfss{D}_N \mathbfss{U}_N^T + \mathbfss{U}_S \mathbfss{U}_S^T,
\end{equation} where 

\begin{equation}
\mathbfss{D}_N = \mathrm{diag} [\lambda_1 + 1, \cdots, \lambda_m +1]
\end{equation} is an $m \times m$ matrix. Therefore,  using Eq. \eqref{eigenstructure} and the orthonormality condition ${\mathbfss{U}_N \mathbfss{U}_N^T + \mathbfss{U}_S \mathbfss{U}_S^T = \mathbfss{I}}$, the foregrounds plus noise covariance matrix can be written as

\begin{align}
\label{estimate}
\widehat{\mathbfss{R}}_n &= \mathbfss{R} - \widehat{\mathbfss{R}}_{\mathrm{HI}}  \nonumber \\ &= \widehat{\mathbfss{R}}_{\mathrm{HI}}^{1/2} (\widehat{\mathbfss{R}}_{\mathrm{HI}}^{-1/2} \mathbfss{R} \widehat{\mathbfss{R}}_{\mathrm{HI}}^{-1/2} - \mathbfss{I}) \widehat{\mathbfss{R}}_{\mathrm{HI}}^{1/2} \nonumber \\ &= \widehat{\mathbfss{R}}_{\mathrm{HI}}^{1/2} (\mathbfss{U}_N (\mathbfss{D}_N - \mathbfss{I}) \mathbfss{U}_N^T ) \widehat{\mathbfss{R}}_{\mathrm{HI}}^{1/2}.
\end{align} Given that the eigenvalues related to foreground components and collected in the diagonal matrix $\mathbfss{D}_N$ are significantly larger than $1$, Eq. \eqref{estimate} can be approximated by

\begin{align}
\label{estimate-approx}
\widehat{\mathbfss{R}}_n &\simeq \widehat{\mathbfss{R}}_{\mathrm{HI}}^{1/2} (\mathbfss{U}_N \mathbfss{D}_N \mathbfss{U}_N^T ) \widehat{\mathbfss{R}}_{\mathrm{HI}}^{1/2}.
\end{align} As the complementary part of the data, the HI covariance matrix can then be represented as the orthogonal subspace of the foregrounds plus noise subspace,

\begin{equation}
\label{estimate1}
\widehat{\mathbfss{R}}_{\mathrm{HI}} = \widehat{\mathbfss{R}}_{\mathrm{HI}}^{1/2} (\mathbfss{U}_S \mathbfss{U}_S^T) \widehat{\mathbfss{R}}_{\mathrm{HI}}^{1/2}.
\end{equation} This is the $n_{\mathrm{ch}} \times n_{\mathrm{ch}}$ HI covariance matrix projected onto the $(n_{\mathrm{ch}} -m)$-dimensional HI subspace spanned by the subset of eigenvectors collected in the matrix $\mathbfss{U}_S$. As long as the HI emission can be effectively described by $n_{\mathrm{ch}} -m$ degrees of freedom, this projection keeps the HI emission and discards the foregrounds plus noise signal from the data. The effective number of degrees of freedom needed to describe the HI emission is limited by the number of available frequency channels, which for HI intensity mapping experiments will be large enough to ensure a safe separation of the HI and the foregrounds plus noise subspaces. 

Using Eq. $\eqref{estimate1}$ and Eq. $\eqref{st}$, we estimate the mixing matrix of the HI signal as the $n_{\mathrm{ch}} \times (n_{\mathrm{ch}} - m)$ matrix
\begin{equation}
\label{mixmat}
\widehat{\mathbfss{S}} = \widehat{\mathbfss{R}}_{\mathrm{HI}}^{1/2} \mathbfss{U}_S.
\end{equation} The estimate of the HI mixing matrix, Eq. \eqref{mixmat}, is the only information needed to implement the multidimensional ILC filter given by Eq. \eqref{multi}. More precisely, the column space $\mathbfss{U}_S$ is the only information needed to use the ILC filter. The exact amplitude of the prior $\widehat{\mathbfss{R}}_{\mathrm{HI}}$ is not critical for the ILC filter because the prior could be multiplied by a constant factor while leaving Eq. \eqref{multi} unchanged. This is true as long as this constant factor is not large enough to modify the dimension of the matrix $\mathbfss{U}_S$. In separating the HI subspace and the foregrounds plus noise subspace, there is a possibility that a subdominant part of foregrounds plus noise signal is projected into the HI subspace. However, the multidimensional ILC minimizes the variance of the foregrounds plus noise signal that may be present in the HI subspace while guaranteeing unit response to the HI signal.



The number $n_{\mathrm{ch}} - m$ of degrees of freedom of the HI emission is expected to vary as we calculate the transformed observation covariance matrix, Eq. \eqref{transcov}, in different areas of the sky. For example, at low Galactic latitude this number may decrease compared to high Galactic latitude because the contribution from Galactic foregrounds to the observed signal becomes larger at this latitude. This number is also expected to vary with angular scale because, unlike foreground emission, the HI signal and the noise have more power on small angular scales. Therefore, to reconstruct accurately the HI emission, we should estimate the dimension of the HI emission subspace locally in space and in angular scale. This is achieved by decomposing the data on a wavelet frame. The use of wavelets (needlets) is described in the Appendix~\ref{app:needlets}.

\subsection{Akaike Information Criterion (AIC)}

To find the local number of degrees of freedom $n_{\mathrm{ch}} - m$ of the HI emission in each wavelet domain, we make use of a statistical information criterion to discriminate between the eigenvalues related to foregrounds and noise and the eigenvalues related to the HI emission, which in our representation (Eq. $\eqref{diag}$) are close to 1. Estimating the dimension $n_{\mathrm{ch}} - m$ of the HI emission subspace is equivalent to counting the effective number $m$ of eigenvalues significantly larger than $1$, which correspond to the foregrounds plus noise degrees of freedom. Instead of determining the number of principal components in an ad-hoc manner as in the standard {\tt PCA}, the effective rank $m$ of the foregrounds plus noise covariance matrix is estimated with the use of the Akaike Information Criterion (AIC) \citep{Akaike1974}.

For each location on the sky and each range of angular scales (wavelet domain), we can select the best rank value $m_b$ for the foregrounds plus noise covariance matrix by minimizing the AIC,

\begin{equation}
\label{aict}
\mathrm{AIC}(m) = 2 \, n m  - 2 \log (\mathcal{L}_{\mathrm{max}}(m)),
\end{equation} where $n$ is the number of modes in the domain considered, $m$ is the number of degrees of freedom of the foregrounds plus noise signal, and $n\,m$ is the number of parameters in the model. $\mathcal{L}_{\mathrm{max}}(m)$ is the maximum likelihood solution of the data covariance matrix given a model of $m$ independent foreground components. For a given number $m$ of independent foreground components, the maximum likelihood solution is given by Eq. \eqref{eigenstructure}.

We note that, as the preferred value $m$ is the one with the minimum AIC value, AIC rewards goodness of fit through the maximum likelihood function, but it also includes a penalty that is an increasing function of the number of dimension of the foregrounds plus noise subspace. This penalty prevent us of overfitting the foreground plus noise subspace. 

In Appendix~\ref{app:aic}, we show that the application of Eq. $\eqref{aict}$ to our problem of finding the dimension of the foreground plus noise subspace in each region considered reduces to the following expression,

 \begin{equation}
\label{aic1t}
\mathrm{AIC}(m) = n \left(2m + \sum_{i=m+1}^{n_{\mathrm{ch}}} [\mu_i - \log \mu_i - 1] \right), 
\end{equation} where $\mu_i$ are the eigenvalues of the transformed covariance matrix of the observed data Eq. \eqref{transcov}.

\subsection{Summary of the GNILC algorithm}

The {\tt GNILC} method presented above can be divided into two main steps. First, using a prior on the HI power spectrum, we determine the local signal to noise ratio and perform a constrained {\tt PCA} of the observed data (section \ref{subsec:cpca}) to determine the effective dimension of the HI subspace. Second, we perform a multimensional ILC filter within the HI subspace (Eq. \eqref{multi}) and reconstruct the HI signal. In the constrained {\tt PCA} step, the number of principal components of the observation covariance matrix is estimated locally both in space and in angular scale by using a wavelet (needlet) decomposition of the observations. We also use a statistical information criterion (AIC) to make the selection of the principal components of the observation covariance matrix (the eigenvectors of the foregrounds plus noise subspace). 

The steps of the {\tt GNILC} method can be summarized as follows: 

\begin{itemize}

\item[1)] To isolate the different ranges of angular scales (wavelet scales), we first define a set of needlet windows in harmonic space (see Appendix \ref{app:needlets}). These needlet windows work as band-pass filters. The spherical harmonic coefficients $a_{\ell m}$ of the observed frequency maps are then band-pass filtered by the needlet windows. Therefore, when we transform back these coefficients into a map, we include statistical information only from a certain range of angular scales. We produce one observed map for each needlet scale $j$.

\item[2)] For each needlet scale $j$, we compute the data covariance matrix, at pixel $p$, of a pair of frequencies $a$ and $b$ as

\begin{equation}
\widehat{\mathbfss{R}}_{ab}(p) = \sum_{p' \in \mathcal{D}(p)}  \textbf{\textit{x}}_a(p') \textbf{\textit{x}}_b^{T}(p'), 
\end{equation} where $\mathcal{D}$ is a domain of pixels centred around the pixel $p$. 

\item[3)] For each needlet scale $j$, we also compute the HI emission covariance matrix by using HI maps $\textbf{\textit{y}}$ simulated from a theoretical prior on the HI angular power spectrum,

\begin{equation}
\widehat{\mathbfss{R}}_{\mathrm{HI} \, ab}(p) = \sum_{p' \in \mathcal{D}(p)}  \textbf{\textit{y}}_a(p') \textbf{\textit{y}}_b^{T}(p').
\end{equation} 

\item[4)] We diagonalize the transformed data covariance matrix, $\widehat{\mathbfss{R}}_{\mathrm{HI}}^{-1/2} \widehat{\mathbfss{R}} \widehat{\mathbfss{R}}_{\mathrm{HI}}^{-1/2}$ as

\begin{equation}
\widehat{\mathbfss{R}}_{\mathrm{HI}}^{-1/2} \widehat{\mathbfss{R}}(m) \widehat{\mathbfss{R}}_{\mathrm{HI}}^{-1/2} = \mathbfss{U}_N \mathbfss{D}_N \mathbfss{U}_N^T + \mathbfss{U}_S \mathbfss{U}_S^T, 
\end{equation} where $\mathbfss{D}_N$ collects the $m$ largest eigenvalues, $\mathbfss{U}_N$ the corresponding eigenvectors, and $\mathbfss{U}_S$ the $(n_{\mathrm{ch}} - m)$ eigenvectors related to the HI emission subspace.

The effective dimension $m$ of the foregrounds plus noise subspace (number of principal components) is estimated by minimizing the AIC,

\begin{equation}
\mathrm{min} \; \left(2m + \sum_{i=m+1}^{n_{\mathrm{ch}}} [\mu_i - \log \mu_i - 1] \right) \;\;\; \mathrm{with} \;\;\; m \in [1, n_{\mathrm{ch}}], \nonumber
\end{equation} where $\mu_i$ are the eigenvalues of $\widehat{\mathbfss{R}}_{\mathrm{HI}}^{-1/2} \widehat{\mathbfss{R}} \widehat{\mathbfss{R}}_{\mathrm{HI}}^{-1/2}$.

\item[5)] For each needlet scale $j$, we apply an $(n_{\mathrm{ch}} - m)$-dimensional ILC filter to the observed data,

\begin{equation}
\widehat{\textbf{\textit{s}}}^{(j)} = \widehat{\mathbfss{S}} (\widehat{\mathbfss{S}}^T \widehat{\mathbfss{R}}^{-1} \widehat{\mathbfss{S}} )^{-1} \widehat{\mathbfss{S}}^T \widehat{\mathbfss{R}}^{-1} \textbf{\textit{x}}^{(j)}, 
\end{equation} where the estimated mixing matrix is given by

\begin{equation}
\widehat{\mathbfss{S}} = \widehat{\mathbfss{R}}_{\mathrm{HI}}^{1/2} \mathbfss{U}_S. 
\end{equation}

\item[6)] Finally, we synthesize the reconstructed needlet HI maps $\widehat{\textbf{\textit{s}}}^{(j)}$ as follows: the maps are transformed to spherical harmonic space, their harmonic coefficients are again band-pass filtered by the respective needlet window and the filtered harmonic coefficients are transformed back to maps in real space. This operation gives one reconstructed HI maps per needlet scale. These maps are then added to give, for each frequency channel, the complete reconstructed HI map.


\end{itemize}

\section{Simulations}
\label{sec:simu}

At radio wavelengths ($\lambda \gtrsim 21$ cm), the astrophysical foregrounds are several orders of magnitude brighter than the HI signal. For example, at $\nu = 1$ GHz, $T_{\mathrm{sky}} \sim 10$ K, while the HI brightness temperature is $T_{\mathrm{HI}} \sim 0.1$ mK \citep{Battye2013}. Here, the sky temperature $T_{\mathrm{sky}}$ is given by the sum of the CMB temperature, the diffuse Galactic radiation, the emission from extragalactic sources, and the HI emission. The CMB emission is not considered in our simulations because it is well constrained by the available data \citep{Planck2015} and consequently it can be easily removed from the observed data.

To test our component separation method we perform three different simulations for the observed sky. The set of simulations is summarized in Table~\ref{tab:simu}. 

\begin{figure}
  	\centering
    \includegraphics[width=0.48\textwidth]{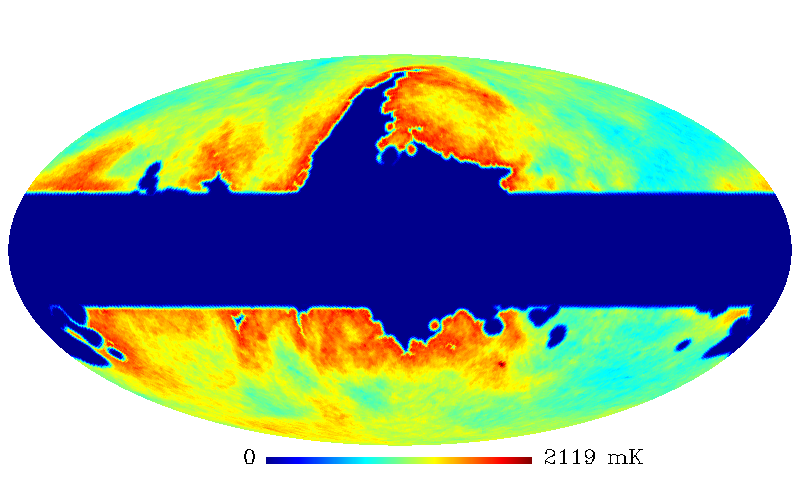}
    \caption{The observed sky at 1117.5 MHz of simulation 1 (HI signal, synchrotron with constant spectral index, and thermal noise) with our Galactic mask.}
    \label{fig:mask}
\end{figure}

\begin{table}
  \footnotesize
   \caption{Simulations for the observed sky of an HI intensity mapping experiment.}
    \label{tab:simu}
    \begin{center}
      \begin{tabular}{| c | p{5cm} |}
        \hline
        Simulation & Components \\ \hline
        1 & HI, synchrotron radiation with constant spectral index, and thermal noise  \\ \hline
        2 & HI, synchrotron radiation with constant spectral index, point sources, free-free radiation, and thermal noise  \\ \hline
        3 & HI, synchrotron radiation with spatially variable spectral index, point sources, free-free radiation, and thermal noise  \\ \hline
      \end{tabular}
    \end{center} 
\end{table}

For each of the simulations, we consider the same bandwidth: 960 to 1260 MHz, which is the proposed range for the BINGO experiment \citep{Battye2013}. This bandwidth corresponds to a redshift range of 0.13 to 0.48. To study the effect of the number of frequency channels in the performance of {\tt GNILC} method, we choose a small number of channels and vary it between 6, 10, and 20. The frequency channels are chosen to be equally spaced in the given bandwidth.

We consider circular Gaussian beams for the frequency channels of the experiment. For simplicity, independently of the number of frequency channels, we fix the full width at half minimum $\theta_{\mathrm{FWHM}}$ of the beam for the first frequency channel to be equal to 40 arcmin, which is the angular resolution of the BINGO telescope at 1 GHz \citep{Battye2013}. Considering that our HI intensity mapping is diffraction-limited, the full width at half minimum of the beam for the other frequency channels can be calculated through the following relation, 

\begin{equation}
\theta_{\mathrm{FWHM}}(\nu) = \theta_{\mathrm{FWHM}}(\nu_0) \frac{\nu_0}{\nu},
\end{equation} where $\nu_0$ corresponds to the first frequency channel. Before applying the {\tt GNILC} method to our simulations, we reconvolve all the maps to the same angular resolution, which we choose to be 40 arcmin.

\begin{figure*}
\centering
\subfigure{%
  \includegraphics[width=0.24\textwidth]{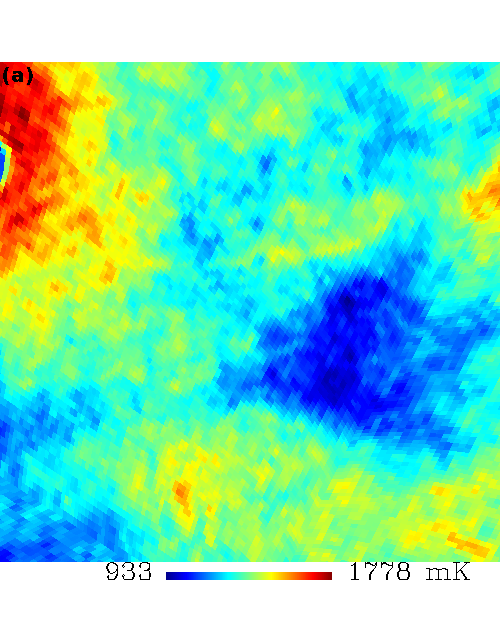}
  \label{fig:synch}}
\subfigure{%
  \includegraphics[width=0.24\textwidth]{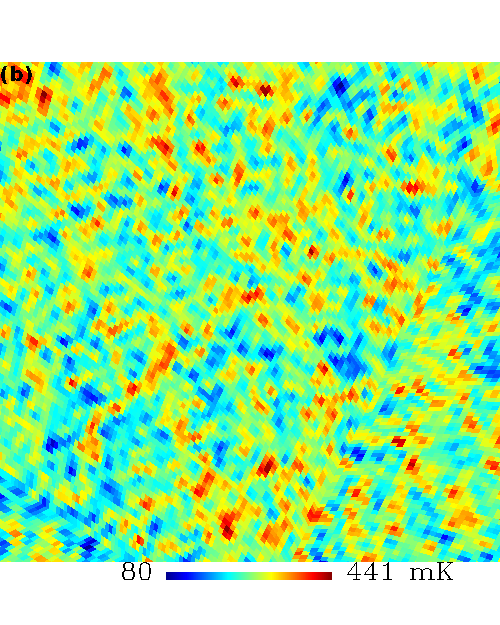}
  \label{fig:ps}}
\subfigure{%
  \includegraphics[width=0.24\textwidth]{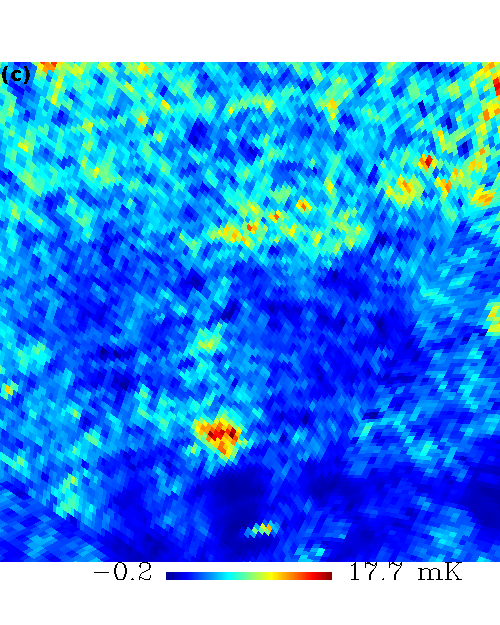}
  \label{fig:ff}}
\subfigure{%
  \includegraphics[width=0.24\textwidth]{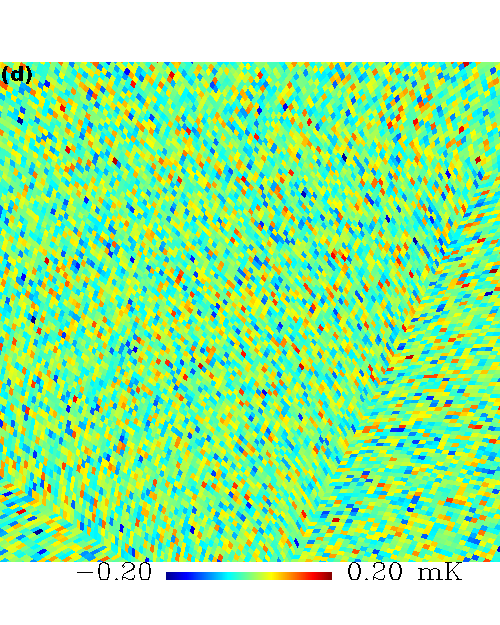}
  \label{fig:noise1}}
\caption{Maps of four simulated signals covering a $50^{\circ} \times 50^{\circ}$ patch of the sky: (a) Galactic synchrotron with constant spectral index, (b) extragalactic point sources, (c) Galactic free-free, (d) and thermal noise. The maps are centred at Galactic coordinates $(120^\circ, - 60^\circ)$ and observed at 1117.5 MHz (redshift $z\sim 0.3$). Note the very different linear scales  used for each map.}
\label{fig:components}
\end{figure*}

We use a Galactic mask to cover the area of the sky where the synchrotron emission of our Galaxy is brightest. Our Galactic mask is given by the product of two masks: a $\pm 20^{\circ}$ latitude mask and a mask designed to cover the Galactic emission with brightness temperatures larger than the threshold of 30 K at 408 MHz. This combined mask is deliberately chosen to allow some moderately bright synchrotron emission to be present in the observed sky; this gives us a stringer test of the {\tt GNILC} method. We also use a cosine apodization of width $3^{\circ}$ in our Galactic mask. This mask gives 51$\%$ of sky coverage. Real experiments typically have smaller areas but this is only an example to study how the {\tt GNILC} method performs with different observational skies and a significantly varying morphology of the foreground emission over the sky. The observed sky of simulation 1 at 1117.5 MHz covered by our Galactic mask is shown in Fig.~\ref{fig:mask}.

To simulate our maps, we use the HEALPix package \citep{Gorski2005} with a resolution $N_{\mathrm{side}} = 128$ and a maximum multipole $\ell_{\mathrm{max}} = 2 \, n_{\mathrm{side}} - 1 = 255$.

Figure~\ref{fig:components} shows some maps of the components of the simulated sky at 1117.5 MHz: synchrotron, free-free, point sources, and the thermal noise. The temperature scale is in mK. The synchrotron radiation temperature is much higher than the temperature of the other foregrounds, roughly one order of magnitude larger than the point sources emission, which is the second dominant foreground. The main contribution from the point sources is in the form of an unresolved background.

\subsection{HI emission}
\label{subsec:hi}

To simulate the HI emission, we use the software CORA \citep{Shaw2014} and assume the cosmological model given in \citet{Planck2014}. We can compute the emitted brightness temperature of the 21 cm line in a velocity width using \citep*{Furlanetto2006}

\begin{equation}
T_o \mathrm{d} v = \frac{\hslash c^3}{k_B} \frac{3 A_{21}}{16 \nu_o^2} n_{\mathrm{HI}} \mathrm{d} l,
\end{equation} where $k_B$ is Boltzmann constant, $T_o$ is the HI brightness temperature in the emitter location, $\nu_o \simeq 1420.406$ MHz is the rest-frame emission frequency for neutral hydrogen, $n_{\mathrm{HI}}$ is the number density of neutral hydrogen atoms, $\mathrm{d}l$ is the line-of-sight distance, and $A_{21}$ is the spontaneous emission coefficient of the 21 cm transition. 

The mean observed brightness temperature due to the average HI density in the Universe is

\begin{equation}
\label{temp}
\bar{T}(z) = \frac{\bar{T}_o(z)}{(1 + z)} = \left( \frac{3 A_{21} \hslash c^3}{16 \nu_o^2 k_B m_H} \right) \left( \frac{\rho_{\mathrm{HI}}(z)}{1 + z} \right) \frac{\mathrm{d} l}{\mathrm{d} v},
\end{equation} where $m_{H}$ is the mass of the hydrogen atom and $\rho_{\mathrm{HI}}(z)$ is the average density of HI at redshift $z$. The first term is a combination of fundamental constants and measured experimental parameters, the second is a function of redshift and the last term relates the line-of-sight distance to the recession velocity.

In the background level, the last term in Eq. $\eqref{temp}$ is given by

\begin{equation}
\frac{\mathrm{d} l}{\mathrm{d} v} = \frac{(1 + z)^3}{H(z)}.
\end{equation} In this way, Eq. $\eqref{temp}$ becomes

\begin{equation}
\label{temp_1}
\bar{T}(z) = \left( \frac{3 A_{21} \hslash c^3}{16 f_o^2 k_B m_H} \right) \frac{(1 + z)^2 \rho_{\mathrm{HI}}(z)}{H(z)}.
\end{equation} 

In a linearly perturbed Universe, we can construct the 2D angular power spectrum of the HI intensity over some frequency range. Assuming that $\mathrm{d} l / \mathrm{d} v$ and $1 + z$ take the values in an unperturbed Universe, which means that we are ignoring the effects of peculiar velocities and the Sachs-Wolfe effect, we can obtain the 3D quantity $\delta \bar{T} (r(z) \widehat{n}, z)$ from Eq. $\eqref{temp_1}$ by replacing $\rho_{\mathrm{HI}}$ with $\delta \rho_{\mathrm{HI}}$. The projection on the sky of the temperature perturbation, $\delta T (\widehat{n})$, is defined, using a window function $W(z)$, which can be taken as uniform in the observed redshift range, in the following way \citep{Battye2013}

\begin{align}
\label{temp_red}
\delta T(\widehat{n}) &= \int \mathrm{d} z \, W(z) \, \delta \bar{T} (r(z) \widehat{n}, z) \nonumber \\ &= \int \mathrm{d} z \, W(z) \, \bar{T}(z) \, \delta_{\mathrm{HI}} (r(z) \widehat{n}, z),
\end{align} where $\delta_{\mathrm{HI}} = \delta \rho_{\mathrm{HI}} / \rho_{\mathrm{HI}}$.

Making a Fourier transform, we have

\begin{align}
\delta \bar{T} (r(z) \widehat{n}, z) &= \bar{T}(z) \int \frac{\mathrm{d}^3 k}{(2 \pi)^3} \widehat{\delta}_{\mathrm{HI}} (\textbf{\textit{k}}, z) e^{i r(z) \widehat{n} \textbf{\textit{k}}} \nonumber \\ &= 4 \pi \bar{T}(z) \sum_{\ell, m} i^{\ell} \int \frac{\mathrm{d}^3 k}{(2 \pi)^3} \widehat{\delta}_{\mathrm{HI}} (\textbf{\textit{k}}, z) j_{\ell} (k r(z)) \nonumber \\  &\qquad \qquad {} \times Y_{\ell m}^*(\widehat{k}) Y_{\ell m} (\widehat{n}), 
\end{align} where $j_{\ell}(x)$ is the spherical Bessel function and $Y_{\ell m}(x)$ are the spherical harmonics. Using this expression in Eq. $\eqref{temp_red}$, we have

\begin{align}
\delta T(\widehat{n}) = 4 \pi &\sum_{\ell, m} i^{\ell} \int \mathrm{d} z W(z) \bar{T}(z) \int \frac{\mathrm{d}^3 k}{(2 \pi)^3} \widehat{\delta}_{\mathrm{HI}} (\textbf{\textit{k}}, z) \nonumber \\  &\qquad {} \times j_{\ell} (k r(z)) Y_{\ell m}^*(\widehat{k}) Y_{\ell m} (\widehat{n}), 
\end{align} which gives the harmonic coefficients

\begin{align}
a_{\ell m} = 4 \pi i^{\ell} \int \mathrm{d} z W(z) \bar{T}(z) \int \frac{\mathrm{d}^3 k}{(2 \pi)^3} \widehat{\delta}_{\mathrm{HI}} (\textbf{\textit{k}}, z) j_{\ell} (k r(z)) Y_{\ell m}^*(\widehat{k}).
\end{align} The angular power spectrum $C_{\ell} $ is defined by the ensemble average $C_{\ell} \equiv \langle a_{\ell m}^* a_{\ell m} \rangle$. Using the orthonormality condition of the spherical harmonics and the definition of the power spectrum,

\begin{equation}
\langle \widehat{\delta}_{\mathrm{HI}} ( \textbf{\textit{k}}, z) \widehat{\delta}_{\mathrm{HI}} ( \textbf{\textit{k}}', z') \rangle = (2 \pi)^3 \delta (\textbf{\textit{k}} - \textbf{\textit{k}}') b^2  P_c (k) D(z) D(z'), 
\end{equation} where $b$ is the bias between the spatial distribution of the HI and the dark matter, $P_c$ is the underlying dark matter power spectrum, and $D(z)$ is the growth factor for dark matter perturbations defined such that $D(0) = 1$, we obtain the angular power spectrum for the HI signal,

\begin{align}
C_{\ell} &= \frac{2 b^2}{\pi} \int \mathrm{d} z W(z) \bar{T}(z) D(z) \int \mathrm{d} z' W(z') \bar{T}(z') D(z') \nonumber \\ &\times \int k^2 \mathrm{d} k P_c(k) j_{\ell} (k r(z)) j_{\ell} (k r(z')).
\end{align} Using small redshift bins, $\Delta z_i$ and $\Delta z_j$, centered in the redshifts $z_i$ and $z_j$, we can write this angular power spectrum for the pair of frequencies $\nu_i$ and $\nu_j$ corresponding to the redshifts $z_i$ and $z_j$ as

\begin{align}
C_{\ell \, i j} &= \frac{2 b^2}{\pi} \int_{\Delta z_i} \mathrm{d} z W(z) \bar{T}(z) D(z) \int_{\Delta z_j} \mathrm{d} z' W(z') \bar{T}(z') D(z') \nonumber \\ &\times \int k^2 \mathrm{d} k P_c(k) j_{\ell} (k r(z)) j_{\ell} (k r(z')).
\end{align} Using this angular power spectrum, the software CORA produces maps of the HI signal with r.m.s fluctuations around 0.1 mK.

\subsection{Thermal noise}

\begin{table}
  \footnotesize
   \caption{Instrumental parameters for a single-dish simulation.}
    \label{tab:instru}
    \begin{center}
      \begin{tabular}{| l | c |}
        \hline
        Parameters &  \\ \hline
        Redshift range [$z_{\mathrm{min}}$, $z_{\mathrm{max}}$] & [0.13, 0.48]  \\ \hline
        Bandwidth [$\nu_{\mathrm{min}}$, $\nu_{\mathrm{max}}$] (MHz) & [960, 1260]  \\ \hline
        Number of feed horns $n_{\mathrm{f}}$ & 80  \\ \hline
        Sky coverage $\Omega_{\mathrm{sur}}$ ($\mathrm{deg}^2$) & 21000 \\ \hline
        Observation time $t_{\mathrm{obs}}$ (yrs) & 1 \\ \hline
        System temperature $T_{\mathrm{sys}}$ (K) & 50 \\ \hline
        Beamwidth at the first channel (arcmin) & 40 \\ \hline
      \end{tabular}
    \end{center} 
\end{table}

For the instrument, we consider only thermal noise, which we assume to respect a uniform Gaussian distribution over the sky. This means that we are not considering $1/f$ noise and other sources of systematic errors. For simplicity, we also take the amplitude of noise to be constant across the frequency channels. The optimal sensitivity per pixel of a single-dish experiment can be defined as follows \citep{Wilson2009}

\begin{equation}
\label{noise}
\sigma_t = \frac{T_{\mathrm{sys}}}{\sqrt{t_{\mathrm{pix}} \Delta \nu}},
\end{equation} where $\Delta \nu$ is the frequency channel width, $T_{\mathrm{sys}}$ is the system temperature, and $t_{\mathrm{pix}}$ is the integration time per pixel defined by

\begin{equation}
t_{\mathrm{pix}} = n_{\mathrm{f}} t_{\mathrm{obs}} \frac{\Omega_{\mathrm{pix}}}{\Omega_{\mathrm{sur}}},
\end{equation} where $n_{\mathrm{f}}$ denotes the number of feed horns, $t_{\mathrm{obs}}$ is the total integration time, $\Omega_{\mathrm{sur}}$ is the survey area, and $\Omega_{\mathrm{pix}} = \theta_{\mathrm{FWHM}}^2$ is the beam area. The values that we use for these parameters are those of the BINGO experiment, except the sky coverage, which is larger and thus gives a larger noise amplitude per pixel. The main parameters of our experiment can be seen in Table~\ref{tab:instru}. 

For the case with 20 frequency channels, the amplitude of noise is equal to 0.05 mK. Later, we also consider the case of this amplitude to be equal to 0.08 mK. Because the total bandwidth of the experiment is fixed, when we decrease the number of frequency channels, we appropriately decrease the amplitude of the noise by using the relation between the amplitude of noise and the frequency channel width given by Eq. $\eqref{noise}$.

\subsection{Synchrotron radiation}

Synchrotron emission arises from energetic charged particles moving in a magnetic field \citep{Rybicki2004}. In our Galaxy, these magnetic fields extend well outside the Galactic plane. For this reason, synchrotron emission is less concentrated in the Galactic plane than free-free radiation. 

The frequency scaling of synchrotron flux emission is in the form of a power law, $I_{\nu} \propto \nu^{\alpha}$. In terms of the Rayleigh-Jeans brightness temperature, we have $T \propto \nu^{\beta}$, with $\beta = (\alpha - 2)$. Typically, at GHz frequencies, $\beta = -2.8 \pm 0.2$, and is variable across the sky \citep{Platania1998, Reich1988, Davies1996}.

For the synchrotron radiation we use the reprocessed Haslam map at 408 MHz \citep{Remazeilles2015}, which includes small-scale fluctuations. We rescale this map to the appropriate frequencies using two models. The first one is a simple power-law that relates the brightness temperature of this radiation to the observed frequency: $T \propto \nu^{-2.8}$. We choose $\beta = - 2.8$ because it is the average value at GHz frequencies as given by \citet{Platania1998}. The other model considers that the synchrotron spectral index is spatially variable. We use the model given by \citet{Miville2008}, which used WMAP intensity and polarization data to do a separation of the Galactic components. In this model the synchrotron spectral index has a mean value of $\beta = -3.0$ and a standard deviation of $\sigma_{\beta} = 0.06$.

\subsection{Free-free radiation}

Free-free emission arises from the interaction of free electrons with ions in ionized media, and is called "free-free" $\,$because of the unbound state of the incoming and outgoing electron \citep{Rybicki2004}. At radio frequencies, this comes from warm ionized gas with typical temperature of $T_e \sim 10^4$ K. 

Theoretical calculations of free-free emission in an electrically medium consisting of ions and electrons give the brightness temperature at frequency $\nu$ as \citep{Dickinson2003}

\begin{equation}
\label{ff}
T_{\mathrm{ff}} \simeq 90 \mathrm{mK} \left( \frac{T_e}{K}\right)^{- 0.35} \left( \frac{\nu}{\mathrm{GHz}}\right)^{\beta} \left(\frac{\mathrm{EM}}{\mathrm{cm}^{-6}\mathrm{pc}} \right), 
\end{equation} where EM is the emission measure representing the integral of the electron density squared along the line-of-sight, ${\mathrm{EM} = \int n_e^2 \mathrm{d} l}$, and $\beta$ is the spectral index, which is around $-2.1$. 

The free-free spectrum is well-defined by a power-law with a temperature spectral index $\beta = -2.1$, which acts to flatten the spectral index of the total continuum of our Galaxy where it has a brightness temperature comparable to that of the synchrotron emission. At intermediate and high Galactic latitudes $(\vert b \vert \gtrsim 10^{\circ})$, the EM can be estimated using H$\alpha$ measurements, which can then be converted to radio brightness temperature assuming that $T_e$ is known and the dust absorption is small. To estimate the EM and obtain the free-free map of our Galaxy, we use the H$\alpha$ map of \citet{Dickinson2003} with $T_e = 7000$ K.

\begin{figure}
  	\centering
    \includegraphics[width=0.5\textwidth]{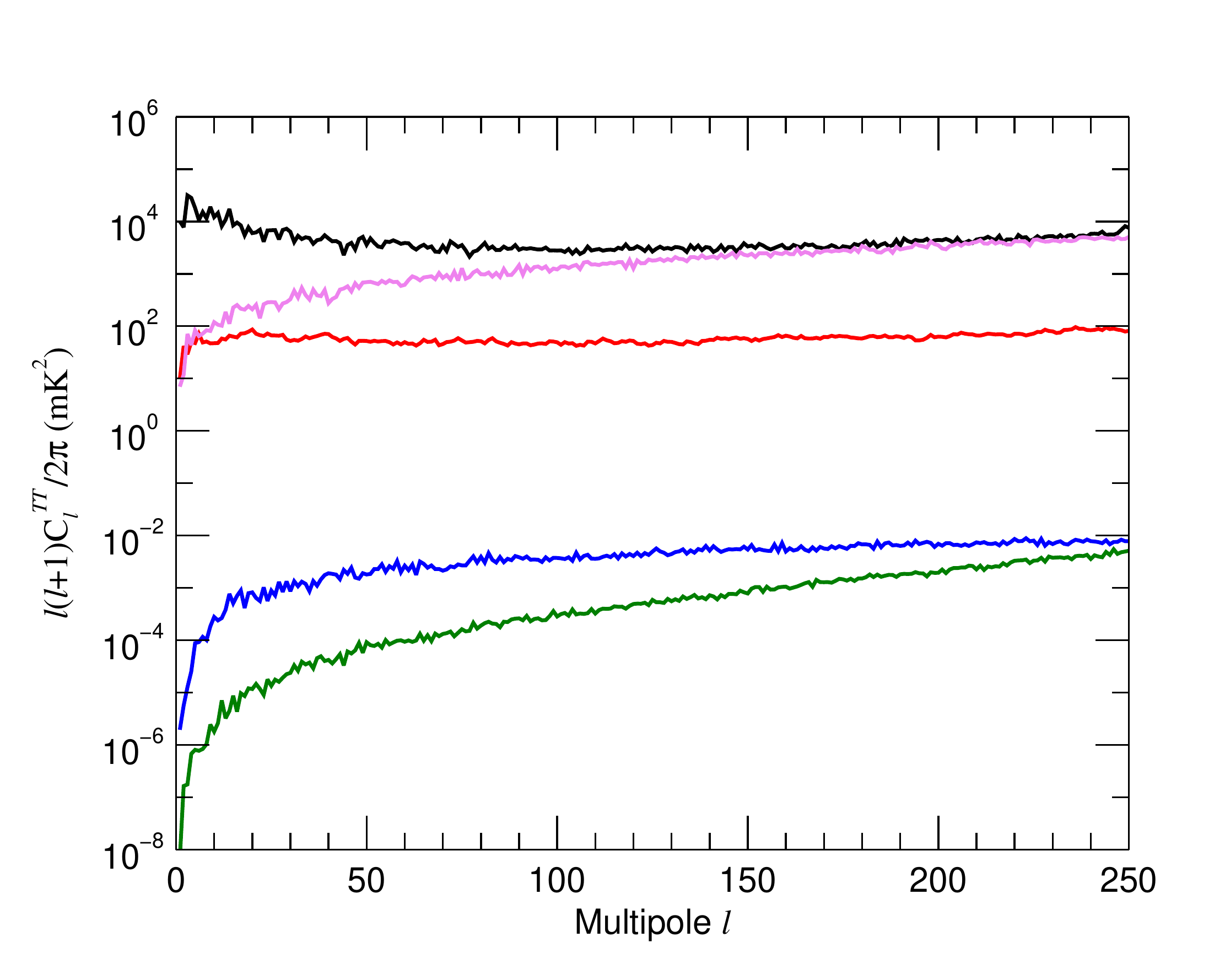}
    \caption{Power spectrum, ${\ell (\ell + 1) C_{\ell} / 2 \pi}$, on a logarithmic scale, of the synchrotron with constant spectral index (black), point sources (pink), free-free (red), HI (blue), and thermal noise (green) at 1117.5 MHz. These power spectra are calculated on 51$\%$ of the sky.}
    \label{fig:ps_comp}
\end{figure}

\subsection{Point sources}

We assume that the distribution of radio point sources is not spatially correlated and hence that they are Poisson distributed \citep{Liu2009}. Extragalactic point sources can be divided into two populations: bright and isolated point sources, which can be detected by the instrument and removed directly
using the observed data, and a continuum of unresolved sources.

We use the model of \citet{Battye2013} to simulate point sources. The brightness of each source is drawn from the differential source counts $\mathrm{d} N / \mathrm{d}S$, where $N$ is the number of sources per steradian and $S$ is the flux.  In \citet{Battye2013}, the authors use data from multiple continuum surveys at 1.4 GHz (see the references in the cited paper) and fit a 5th order polynomial to these data,

\begin{equation}
\log_{10} \left( \frac{S^{2.5} \mathrm{d}N / \mathrm{d}S}{N_0} \right) = \sum_{i=0}^ 5 a_i \left[ \log_{10} \left(\frac{S}{S_0} \right) \right]^i, 
\end{equation} where $a_0 = 2.593$, $a_1 = 9.333 \times 10^{-2}$, $a_2 = - 4.839 \times 10^{-4}$, $a_3 = 2.488 \times 10^{-1}$, $a_4 = 8.995 \times 10^{-2}$, and $a_5 = 8.506 \times 10^{-3}$. We also have $N_0 = 1 \; \mathrm{Jy}^{3/2} \mathrm{sr}^{-1}$ and $S_0 = 1$ Jy.  

The spectral distribution of the sources is given by a power-law function,

\begin{equation}
S(\nu) = S(1.4 \; \mathrm{GHz}) \left(\frac{\nu}{1.4 \; \mathrm{GHz}} \right)^{\alpha},
\end{equation} where the spectral index $\alpha$ is randomly chosen from a Gaussian distribution,

\begin{equation}
P(\alpha) = \frac{1}{\sqrt{2 \pi \sigma^2}} \exp \left[ \frac{(\alpha - \alpha_0)^2}{2 \sigma^2} \right],
\end{equation} with a mean equals to $\alpha_0 = - 2.5$ and a variance equals to $\sigma = 0.5$ \citep{Tegmark2000}.

 \begin{figure*}
\centering
\subfigure{%
  \includegraphics[width=0.3\textwidth]{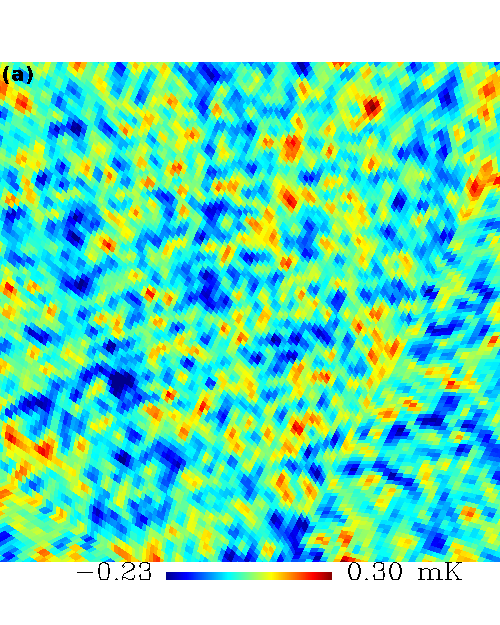}
  \label{fig:i21}}
\subfigure{%
  \includegraphics[width=0.3\textwidth]{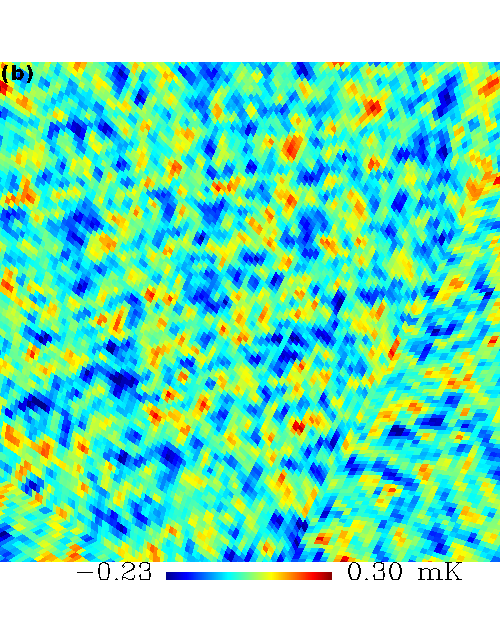}
  \label{fig:o21}}
\quad
\subfigure{%
  \includegraphics[width=0.3\textwidth]{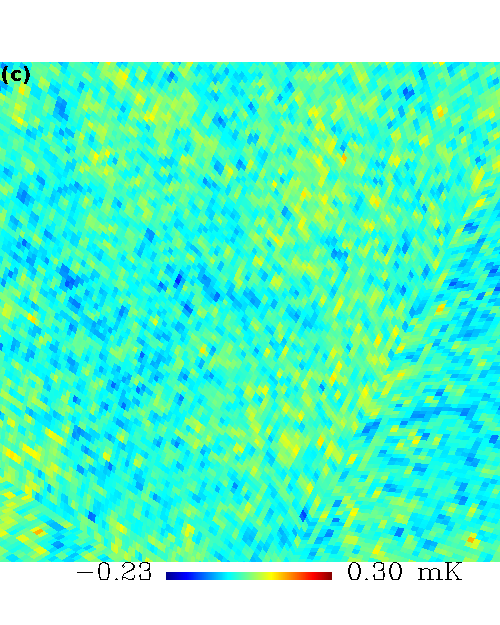}
  \label{fig:residual}}
\caption{Maps covering a $50^{\circ} \times 50^{\circ}$ patch of the sky with three different signals: (a) input HI, (b) {\tt GNILC} reconstructed HI, and (c) residuals. The maps are centred at Galactic coordinates $(120^\circ, - 60^\circ)$ and observed at 1117.5 MHz (redshift $z\sim 0.3$).}
\label{fig:21cm}
\end{figure*}

Assuming that the sources with flux $S > S_{\mathrm{max}}$ can be subtracted from the data, the brightness temperature is given, for each pixel $p$, by the sum of all the point sources contained on it,

\begin{equation}
T_{\mathrm{ps}}(\nu, p) = \left(\frac{\mathrm{d}B}{\mathrm{d}T} \right)^{-1} \, \Omega_{\mathrm{sky}}^{-1} \sum_{i=1}^N S_i^* \left(\frac{\nu}{1.4 \; \mathrm{GHz}} \right)^{\alpha},
\end{equation} where $S_i^*$ is the flux of the point source $i$ at 1.4 GHz, $N$ is the number of point sources in pixel $p$, $\Omega_{\mathrm{sky}} = \theta_{\mathrm{FWHM}}^2$ with $\theta_{\mathrm{FWHM}}$ being the full width at half minimum of the beam of the map at frequency $\nu$, and $\mathrm{d}B/\mathrm{d}T = 2 k_B /\lambda^2$ is the conversion factor between intensity and brightness temperature, with $k_B$ being the Boltzmann constant and $\lambda$ the wavelength of the observed radiation. For our simulation, we assume $S_{\mathrm{max}} = 100$ mJy. 

\section{Results}
\label{sec:results}

\subsection{Basic results}

The angular power spectra of the simulated astrophysical foregrounds, HI emission, and thermal noise are shown in Fig.~\ref{fig:ps_comp}. These power spectra are computed with the HEALPix routine {\sc anafast} \citep{Gorski2005} on the sky  covered with the apodized Galactic mask of Fig.~\ref{fig:mask}. We see that the synchrotron radiation is dominant on all scales and has more power on large angular scales ($\ell < 50$). The extragalactic point sources have the opposite behavior, with its main contribution coming from high multipoles $\ell > 100$. The free-free radiation, on the other hand, has an almost constant power spectrum over the range of scales ($0 < \ell < 250$) and the fraction of the sky considered. Finally, the thermal noise and the HI emission have more power on small scales ($\ell > 150$).

We now present the results of the application of the {\tt GNILC} method to separate the HI signal from the astrophysical and instrumental contamination of a general HI intensity mapping experiment. Figure~\ref{fig:21cm} shows, for simulation 1 (synchrotron with constant spectral index, HI, and thermal noise only) with 20 frequency channels, the input HI signal, the reconstructed HI signal and the residual contamination (recovered HI signal minus input HI signal) at 1117.5 MHz. These maps are shown with an angular resolution of 40 arcmin. We see that, after cleaning foregrounds and thermal noise of the observed signal, the {\tt GNILC} method is able to reconstruct the temperature fluctuations of the HI signal with a r.m.s residual equals to 0.04 mK at this $50^{\circ} \times 50^{\circ}$ patch of the sky. There remains some low-level large-scale residuals, which are also visibile in the power spectra on the largest angular scales ($\ell \lesssim 30$).

\begin{figure*}
  	\centering
    \includegraphics[width=0.95\textwidth]{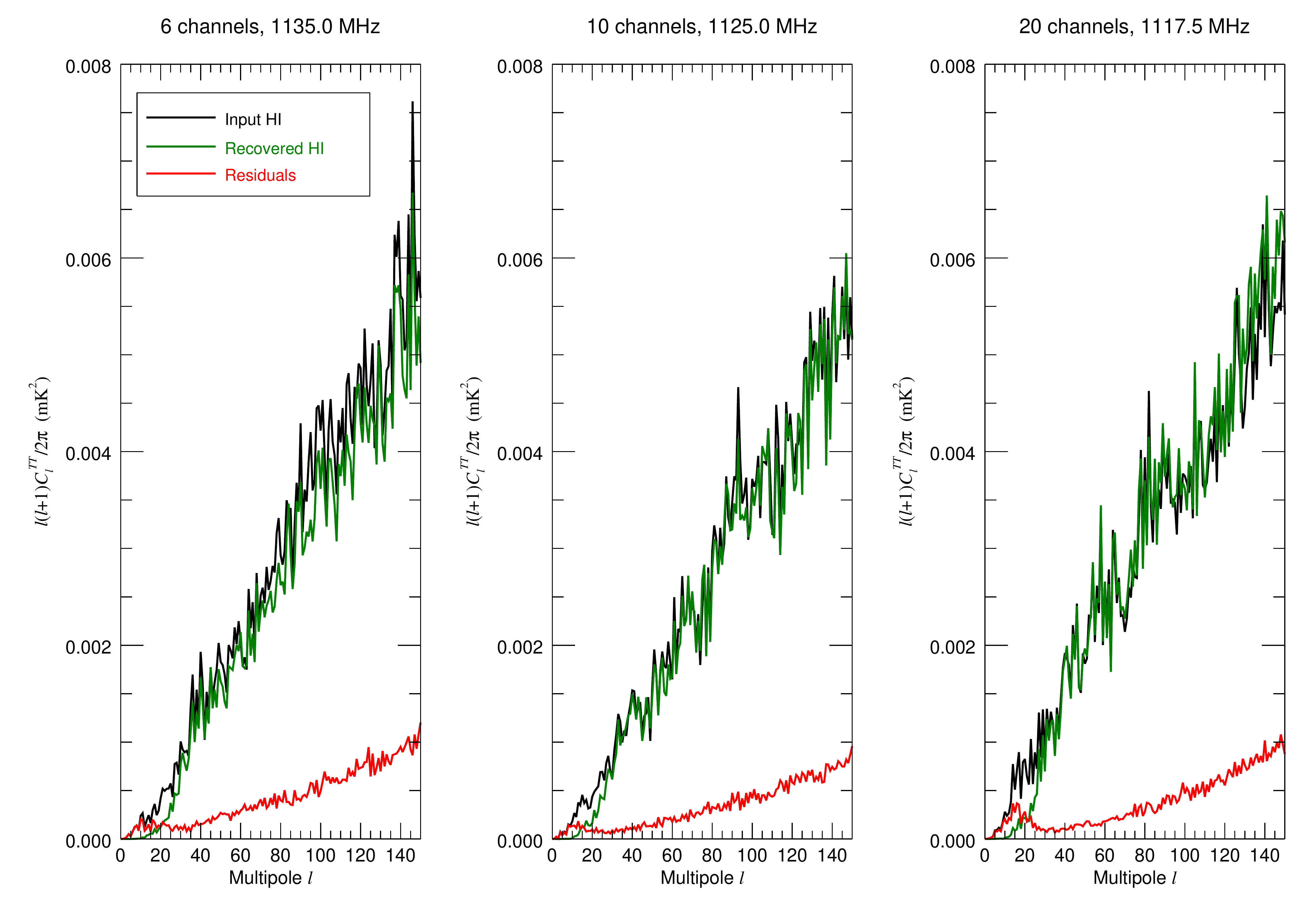}
    \caption{Results on simulation 1 (HI, synchrotron with constant spectral index, and thermal noise) with 6, 10, and 20 frequency channels: power spectra, ${\ell (\ell + 1) C_{\ell} / 2 \pi}$, of the simulated HI signal (black), the recovered HI signal (green), and the residual signal (red) at frequency 1135.0 MHz, 1125.0 MHz, and 1117.5 MHz, respectively.}
    \label{fig:s1a}
\end{figure*}

To have a quantitative measure of the ability of the {\tt GNILC} method to recover the HI signal, we plot, in Fig.~\ref{fig:s1a}, the power spectra of the input HI signal, the reconstructed HI signal, and the residual map on a linear scale. 

Although we have one reconstructed HI map for each frequency, to summarize our results, we show the power spectrum for the frequency that is closest to the middle point of the band. Given that the band extends from 960 MHz to 1260 MHz, for the case of 6, 10, and 20 frequency channels, the central frequency channel is equal to 1135 MHz, 1125 MHz, and 1117.5 MHz, respectively. 

Figure~\ref{fig:s1a} shows that the {\tt GNILC} method reconstructs, without significant bias, the input HI power spectrum for $30 < \ell <150$. In the case of 20 frequency channels, the residual map power spectrum is around 12.5$\%$ of the input HI power spectrum for a range of multipoles that goes from 30 to 150. Also, for the three different numbers of frequency channels, the local features of the HI power spectrum are reconstructed with good accuracy.

When we increase the number of frequency channels, we see that the {\tt GNILC} method performance improves. This improvement with the number of frequency channels is clearer when we plot the normalized difference between the input power spectrum, $C_{\ell}^S$, and the reconstructed power spectrum, $C_{\ell}^R$, of the HI signal,

\begin{equation}
\label{norma}
\Delta C_{\ell} = \frac{C_{\ell}^S - C_{\ell}^R}{C_{\ell}^S}.
\end{equation}

\begin{figure}
  	\centering
    \includegraphics[width=0.5\textwidth]{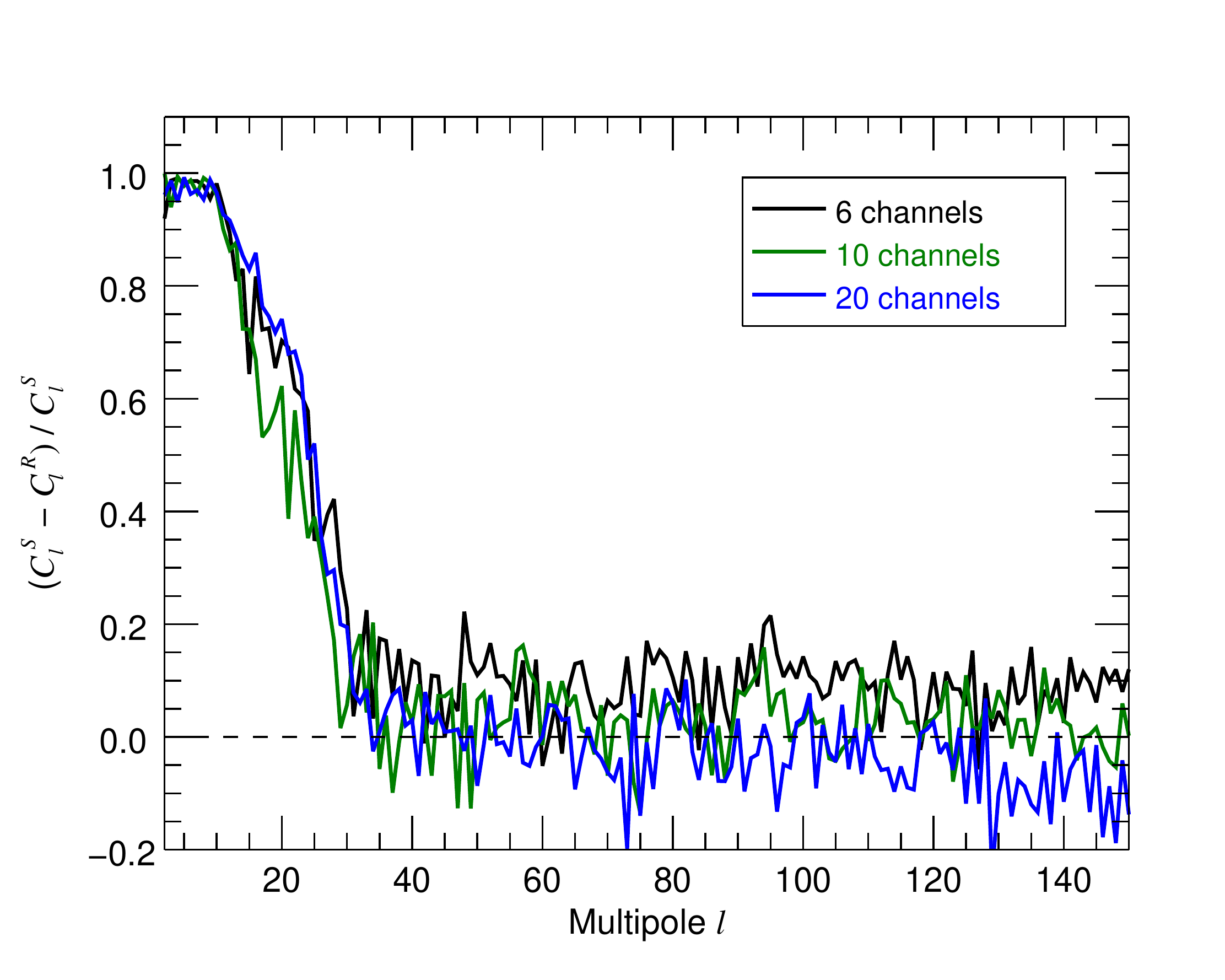}
    \caption{Results on simulation 1 (HI, synchrotron with constant spectral index, and thermal noise): normalized difference between the input power spectrum, $C_{\ell}^S$, and the reconstructed power spectrum, $C_{\ell}^R$, of the HI signal at frequency 1135.0 MHz, 1125.0 MHz, and 1117.5 MHz with 6 (black), 10 (green), and 20 (blue) frequency channels, respectively.}
    \label{fig:s1r}
\end{figure}

\begin{figure*}
  	\centering
    \includegraphics[width=0.95\textwidth]{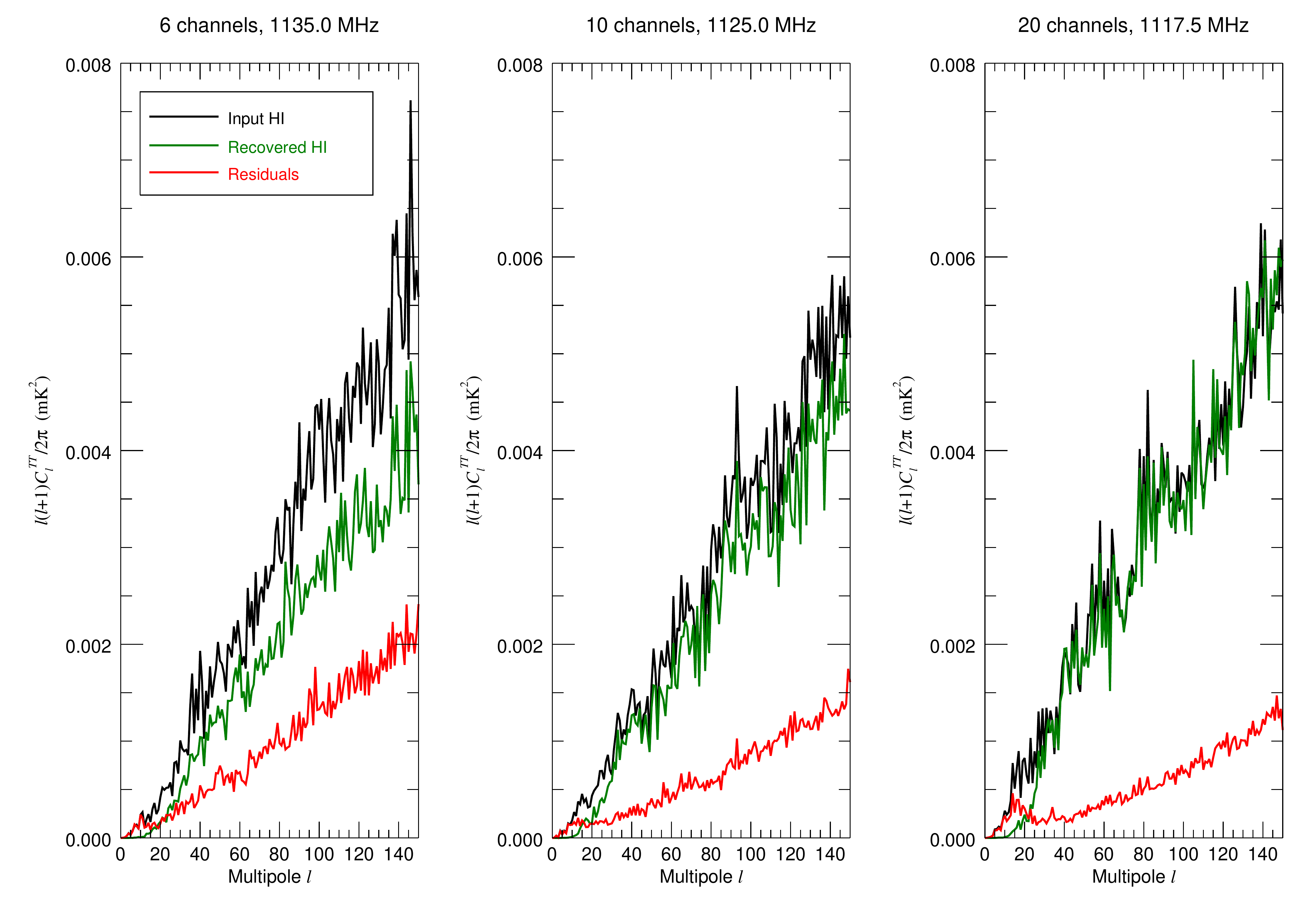}
    \caption{Results on simulation 2 (HI, synchrotron with constant spectral index, point sources, free-free, and thermal noise) with 6, 10, and 20 frequency channels: power spectra, ${\ell (\ell + 1) C_{\ell} / 2 \pi}$, of the simulated HI signal (black), the recovered HI signal (green), and residual signal (red) at frequency 1135.0 MHz, 1125.0 MHz, and 1117.5 MHz, respectively.}
    \label{fig:s2a}
\end{figure*}

This plot is given in Fig.~\ref{fig:s1r}. We see that over a multipole range $30 < \ell < 120$, the HI power spectrum is better reconstructed by the {\tt GNILC} method with 10 and 20 frequency channels than with 6 channels. In this range of multipoles, the average normalized absolute difference between the input power spectrum and the reconstructed power spectrum of the HI signal is 10.9$\%$, 6.0$\%$, and 4.9$\%$ for 6, 10, and 20 frequency channels, respectively. The improvement of the {\tt GNILC} method with the increase of the number of frequency channels can be justified with the help of Eq. $\eqref{diag}$. With more frequency channels, there is more freedom for the {\tt GNILC} method to fit for the independent components of emission because now the effective number of independent degrees of freedom, $m$, required to properly describe the foregrounds plus noise signal is no longer limited by a small number of observations. Therefore, with more frequency channels, we are able to remove foregrounds from the observed data more efficiently.

For multipoles greater than 120, the reconstruction with 20 frequency channels has a greater dispersion (in this case, the average normalized absolute difference between the input power spectrum and the reconstructed power spectrum of the HI signal is equal to 8.9$\%$) than the reconstruction with 10 frequency channels (average normalized absolute difference is equal to 4.2$\%$). This happens because the experiment with 20 frequency channels has a larger level of instrumental noise in the observed data than the experiment with 10 frequency channels (see Eq. $\eqref{noise}$). 
Thus a direct comparison between the two experiments is irrelevant on small scales. 

\begin{figure}
  	\centering
    \includegraphics[width=0.50\textwidth]{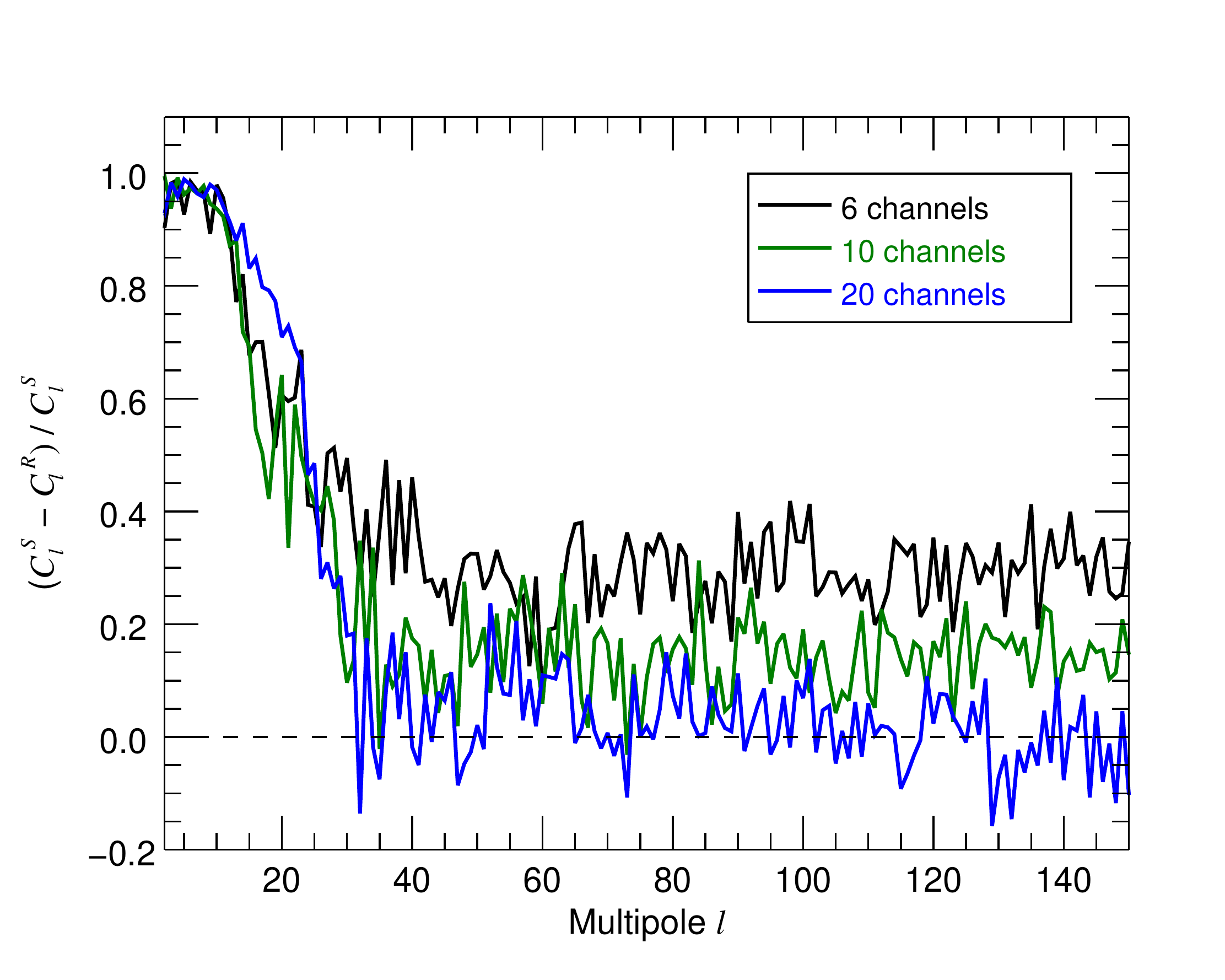}
    \caption{Results on simulation 2 (HI, synchrotron with constant spectral index, point sources, free-free, and thermal noise): normalized difference between the input power spectrum, $C_{\ell}^S$, and the reconstructed power spectrum, $C_{\ell}^R$, of the HI signal at frequency 1135.0 MHz, 1125.0 MHz, and 1117.5 MHz with 6 (black), 10 (green), and 20 (blue) frequency channels, respectively.}
    \label{fig:s2r}
\end{figure}

For simulation 2, we add radio point sources and free-free radiation to simulation 1. This increases the complexity of the observed sky. In this case, the foregrounds plus noise signal has more degrees of freedom than in the case of simulation 1. Figure~\ref{fig:s2a} shows the power spectra of the recovered HI signal, the input HI signal, and the residual signal for simulation 2 with 6, 10, and 20 frequency channels.


As for the case in simulation 1, when we increase the number of frequency channels for simulation 2, the {\tt GNILC} method performance improves, now with a more noticeable effect. For a range of multipoles that goes from 30 to 150, the residual map power spectrum is around 34.7$\%$, 24.3$\%$, and 18.7$\%$ of the input HI power spectrum for the case with 6, 10, and 20 frequency channels, respectively. This improvement of the {\tt GNILC} method with the increase of the number of frequency channels can again be justified with the help of Eq. $\eqref{diag}$. As the presence of point sources and free-free increases the number of degrees of freedom of the foregrounds plus noise signal, we need more frequency channels, $n_{\mathrm{ch}}$, to be able to remove efficiently these extra degrees of freedom from the observed signal.


\begin{figure}
  	\centering
    \includegraphics[width=0.5\textwidth]{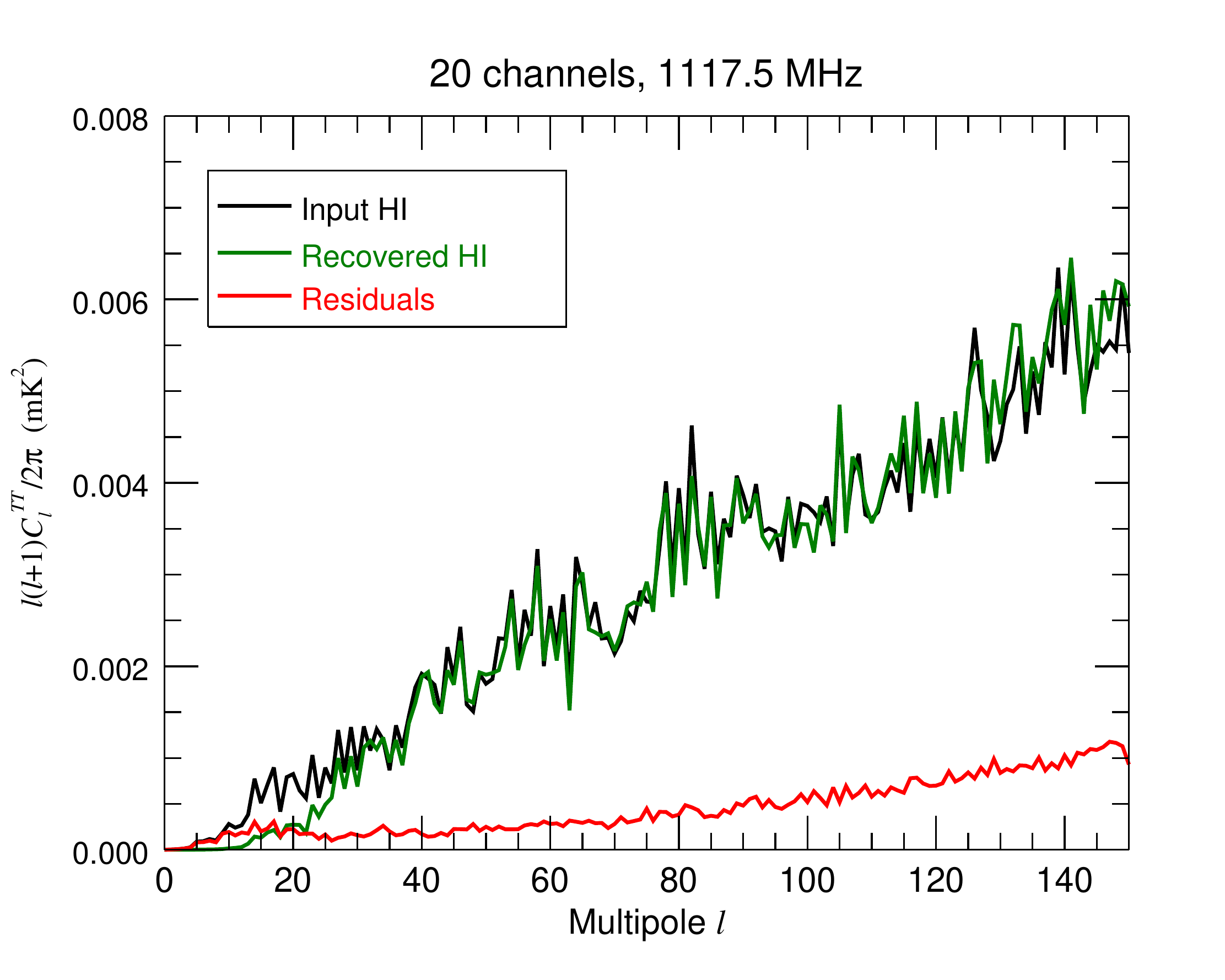}
    \caption{Result on simulation 3 (HI, synchrotron with spatially variable spectral index, point sources, free-free, and thermal noise): power spectra, ${\ell (\ell + 1) C_{\ell} / 2 \pi}$, of the simulated HI signal (black), the recovered HI signal (green), and the residual signal (red) at frequency 1117.5 MHz with 20 frequency channels.}
    \label{fig:s3}
\end{figure}

Figure~\ref{fig:s2r} shows the normalized difference between the input power spectrum, $C_{\ell}^S$, and the reconstructed power spectrum, $C_{\ell}^R$, of the HI signal for simulation 2. For the case of 20 frequency channels, we obtain an average normalized absolute difference around 6.3$\%$ for multipoles in the range between 30 and 150.

In simulation 3, we consider the observed sky of simulation 2 but make the synchrotron radiation spectral index to be spatially variable. The results for this simulation are consistent with those of simulation 2. For the case of 20 frequency channels, we obtain an average normalized absolute difference between the input power spectrum, $C_{\ell}^S$, and the reconstructed power spectrum, $C_{\ell}^R$, of the HI signal that is around 6.1$\%$ for multipoles in the range between 30 and 150. Figure~\ref{fig:s3} shows the power spectra of the recovered HI signal, the input HI signal, and the residual map for simulation 3 with 20 frequency channels.

In Table~\ref{tab:simu_r}, we summarize the values of the average normalized absolute difference between the input power spectrum, $C_{\ell}^S$, and the reconstructed power spectrum, $C_{\ell}^R$, of the HI signal for our simulations.

In the case where the observed sky of a real experiment is more complex than the sky of simulation 3, the number of degrees of freedom required to describe the foregrounds plus noise signal may be larger. Consequently, to reconstruct accurately the HI power spectrum, the {\tt GNILC} method may require more than 20 frequency channels. As HI intensity mapping experiments such as BINGO can have $\sim 1000$ or more channels, the increase, for a real experiment, in the number of degrees of freedom of the foregrounds plus noise signal will not be a limitation for the {\tt GNILC} method.

\subsection{Instrumental effects}

The thermal noise amplitude, for 20 frequency channels, $\sigma_t = 0.05$ mK, chosen for our simulations is of the correct order of magnitude for a single-dish experiment such as BINGO \citep{Bigot-Sazy2015}. To see the effect of a larger amplitude for the thermal noise, we simulate again the observed sky of simulation 1 (HI, synchrotron with constant spectral index, and thermal noise) but now with a thermal noise amplitude $\sigma_t = 0.08$ mK. 

\begin{table}
  \footnotesize
  \scriptsize 
   \caption{The average absolute difference between the input and the {\tt GNILC} reconstructed HI power spectrum normalized by the input HI power spectrum for multipoles in the range between 30 and 150.}
    \label{tab:simu_r}
    \begin{center}
      \begin{tabular}{| c | c | c | c |}
        \hline
        Number of channels & Simulation 1 &  Simulation 2 & Simulation 3 \\ \hline
        6 & 0.11 & 0.30  & 0.27 \\ \hline
        10 & 0.06 & 0.15 & 0.14 \\ \hline
        20 & 0.06 & 0.06 & 0.06 \\ \hline
      \end{tabular}
    \end{center} 
\end{table}

\begin{figure*}
  	\centering
    \includegraphics[width=0.95\textwidth]{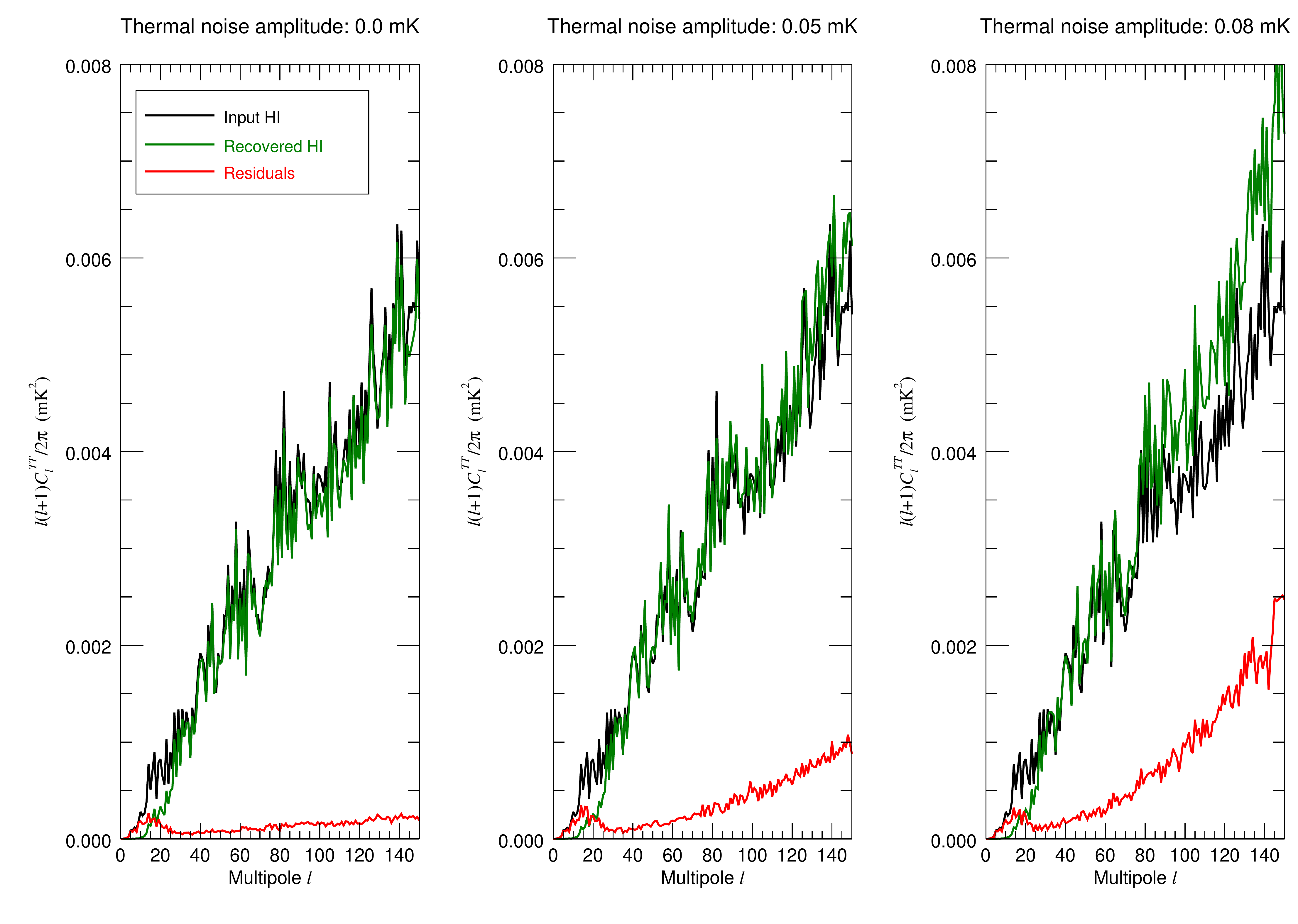}
    \caption{Impact of thermal noise: power spectra, ${\ell (\ell + 1) C_{\ell} / 2 \pi}$, of the simulated HI signal (black), the recovered HI signal (green), and the residual signal (red) for simulation 1 (HI, synchrotron with constant spectral index, and thermal noise) at frequency 1117.5 MHz for different thermal noise amplitudes.}
    \label{fig:noise}
\end{figure*}

In Fig.~\ref{fig:noise}, we show the power spectra of the recovered HI signal, the input HI signal, and the residual map for different noise levels at 1117.5 MHz. In this figure, we have the results for simulation 1 with no thermal noise and with thermal noise amplitudes $\sigma_t$ equal to 0.05 and 0.08 mK.  

We see that some residual thermal noise signal contaminates the recovered HI signal. This effect is significant at small scales ($\ell > 80$), where the recovered HI signal is enhanced, compared to the case without thermal noise, in the cases with thermal noise amplitudes equal to 0.05 and 0.08 mK. As expected, the larger the thermal noise amplitude is, the less accurate is the reconstruction of the HI power spectrum on small angular scales.

However, in a real experiment, if we are able to have a good estimate of the thermal noise power spectrum, then the {\tt GNILC} method is able to recover the HI plus the thermal noise signal from the observed data. To do this, instead of using the HI covariance matrix, $\widehat{\mathbfss{R}}_{\mathrm{HI}}$, in the transformation of Eq. $\eqref{white}$, we can use $\widehat{\mathbfss{R}}_{\mathrm{HI}} + \widehat{\mathbfss{R}}_{\mathrm{noise}}$ as a prior. In this case, we are able to explore the HI signal plus noise subspace and therefore to recover the HI signal plus thermal noise. In this scenario, to minimize the noise bias in the reconstructed power spectrum, we can correct for the thermal noise afterwards by using the estimate of the instrumental noise power spectrum in each frequency channel and then obtain a reconstructed HI power spectrum with minimal noise contamination.

We also consider the possibility that the telescope creates systematic errors that result in non-ideal calibration of the data. To simulate this in a simplified manner, we added a small source of fluctuations in the spectral response. These fluctuations might represent standing waves in the telescope structure or frequency variations in the receiver gain temperature. We parametrize this fluctuations with an additional oscillatory term to the measured brightness temperature as a function of frequency,
\begin{equation}
\label{sin}
\Delta T_b(\nu) = A \sin \left( \frac{\pi \nu}{\Delta \nu_{\mathrm{osc}}} \right),
\end{equation} 
where $A = 100$ mK, $\nu$ is the frequency of observation and $\Delta \nu_{\mathrm{osc}} = $ 10, 100, and 300 MHz. When $\Delta \nu_{\mathrm{osc}}$ is larger than the frequency bin of the experiment ($15$\,MHz for 20 frequency channels in our experiment), we have a curvature in the frequency spectrum, which is similar to the effect of standing waves in a real experiment. On the other hand, when $\Delta \nu_{\mathrm{osc}}$ is smaller than the frequency channel width, it adds a noise-like term, similar to the HI signal itself. This could represent random bandpass calibration errors, for example.

\begin{figure}
  	\centering
    \includegraphics[width=0.49\textwidth]{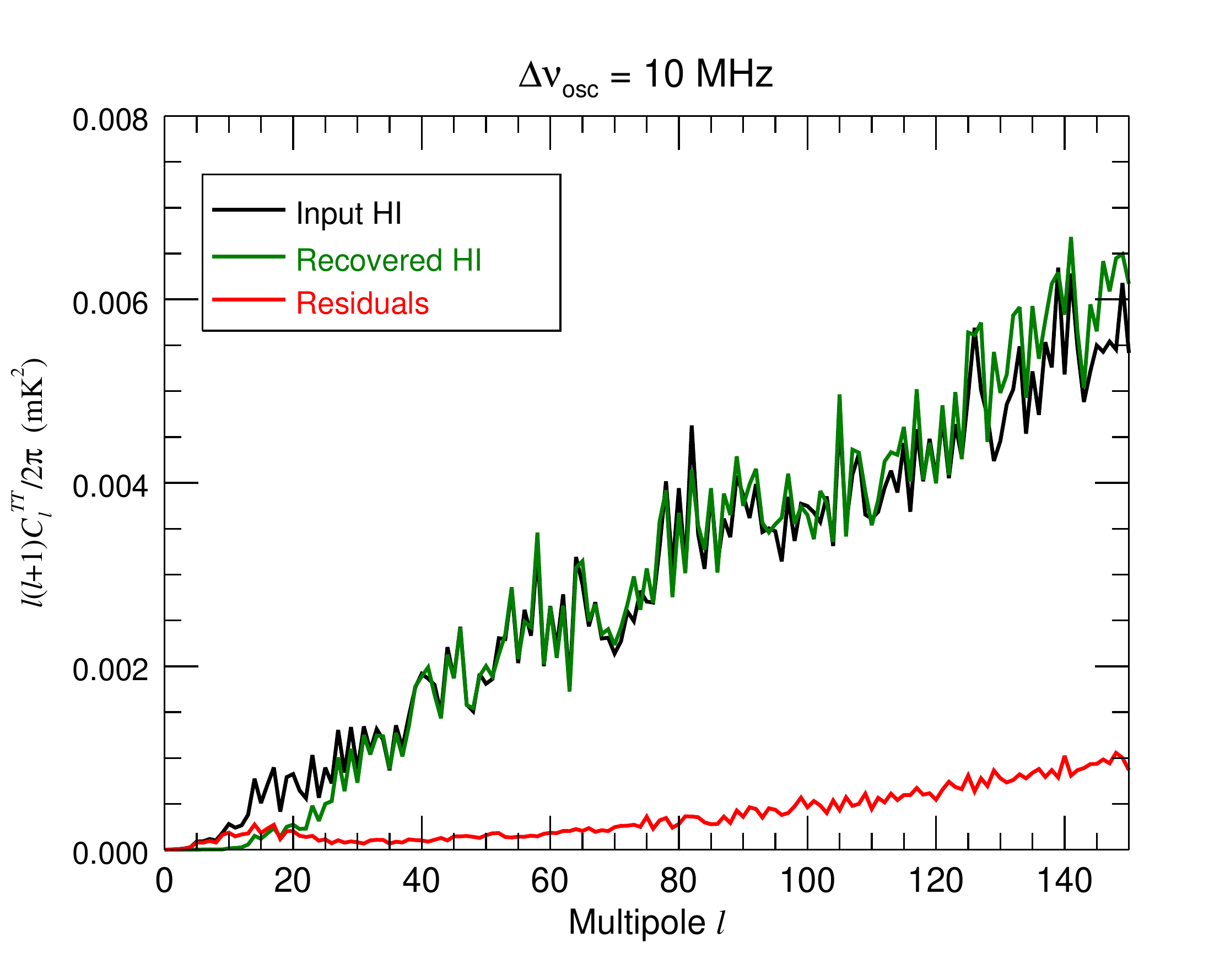}
    \caption{Impact of non-smooth response on frequency: power spectra, ${\ell (\ell + 1) C_{\ell} / 2 \pi}$, of the simulated HI signal (black), the recovered HI signal (green), and the residual signal (red) for the simulation with non-smooth response of the telescope to the synchrotron radiation with $\Delta \nu_{\mathrm{osc}} = 10$ MHz. We use the observed sky of simulation 1 (HI, synchrotron with constant spectral index, and thermal noise).}
    \label{fig:s3a}
\end{figure}

For a range of multipoles that goes from 30 to 150, the average absolute difference between the input and the reconstructed HI power spectrum normalized by the input HI power spectrum is 5.95$\%$,  5.99$\%$, and  5.93$\%$ for the case with $\Delta \nu_{\mathrm{osc}}$ equals to 10, 100, and 300 MHz, respectively. In Fig.~\ref{fig:s3a}, we show the result for $\Delta \nu_{\mathrm{osc}}$ = 10 MHz. We see that the {\tt GNILC} method is robust to the possibility of the telescope having a non-smooth response to the synchrotron radiation frequency spectrum. This shows that the {\tt GNILC} method is able to detect the extra degrees of freedom in the foregrounds plus noise subspace given by the parameters $A$ and $\Delta \nu_{\mathrm{osc}}$ and properly describe this subspace. In general, as long as the angular power spectrum of the systematic effects behaves differently from the angular power spectrum of the HI emission, the {\tt GNILC} method is able to detect the degrees of freedom related to the different systematic effects that are usually present in an HI intensity mapping experiment.

\subsection{Prior effects}
\label{subsec:prior}

The {\tt GNILC} method makes use of a prior of the theoretical HI power spectrum to estimate the dimension of the HI subspace. In this section, we study how the {\tt GNILC} method performs with different priors for the HI power spectrum. First, we use a prior for the HI power spectrum that does not contain any fluctuation, i.e a smooth HI power spectrum, but that has the correct average amplitude for the HI signal. We also study the effect of having a smooth prior that over or underestimate the HI signal. Moreover, we neglect in the prior any possible cross-correlations between different frequency channels (cross power spectra).

\begin{figure*}
  	\centering
    \includegraphics[width=0.95\textwidth]{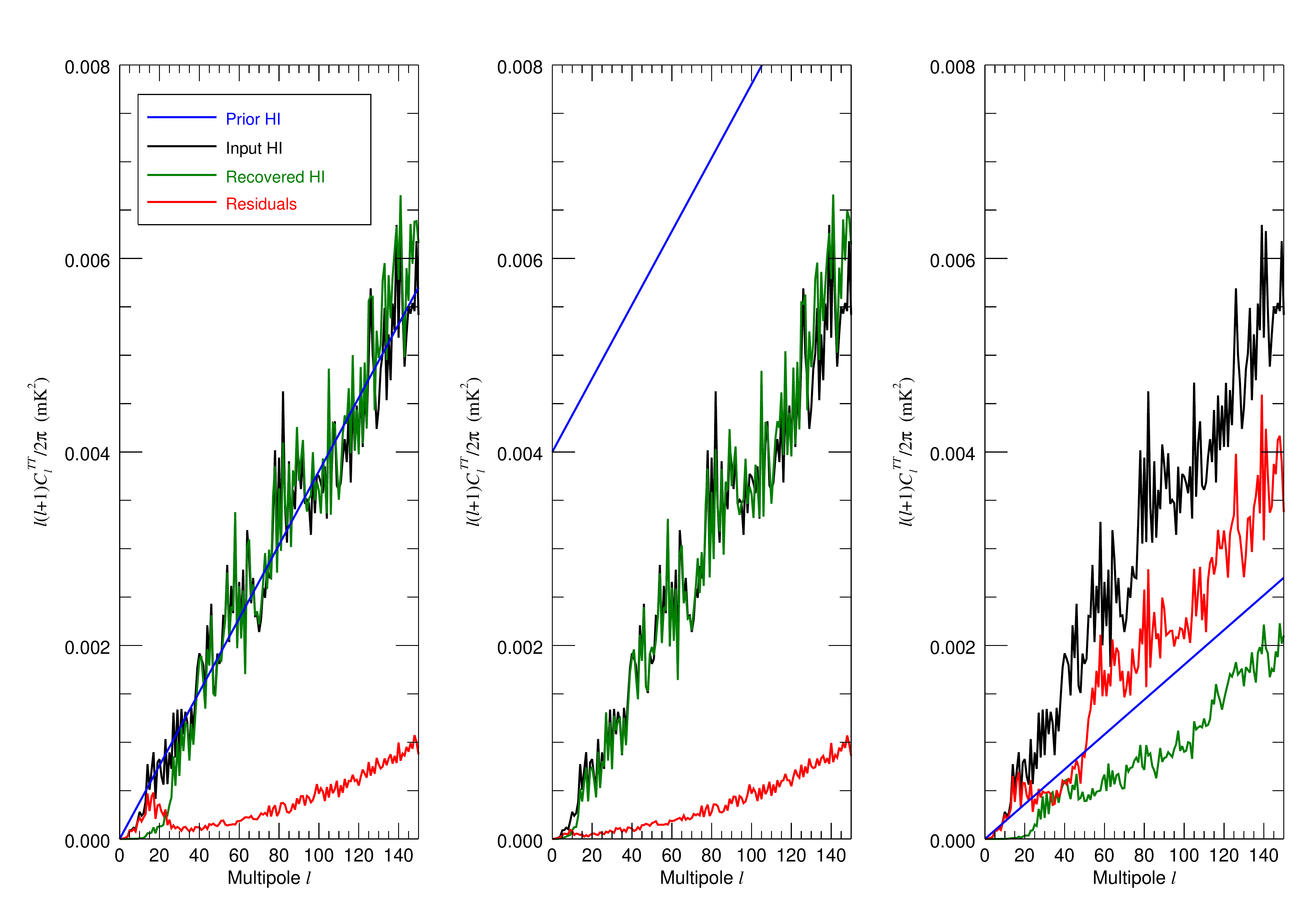}
    \caption{Sensitivity to incorrect priors: HI prior (blue), input HI signal (black), recovered HI signal (green), and residual signal (red) at frequency 1117.5 MHz for the simulation 1 (HI, synchrotron with constant spectral index, and thermal noise) with 20 frequency channels when considering different priors for the HI power spectrum. The prior in the first panel is given by a smooth HI power spectrum with the correct average amplitude for the HI signal, the prior in the second panel is given by a smooth spectrum that overestimates the HI signal by a factor of 2.5, and the prior in the third panel is given by a smooth spectrum that underestimates the HI signal by a factor of 2.}
    \label{fig:s1t}
\end{figure*}

Three examples of these incorrect priors for the HI power spectrum are shown in Fig.~\ref{fig:s1t}. In these simulations, we consider the observed sky of simulation 1 and 20 frequency channels. Figure~\ref{fig:s1t} shows, for each prior considered, the input HI power spectrum, the prior HI power spectrum, the reconstructed HI power spectrum, and the power spectrum of the residual signal.

We see that the {\tt GNILC} method is not critically sensitive to the absence of local features in the prior HI power spectrum. The {\tt GNILC} reconstruction, however, depends on the overall amplitude of the HI power spectrum. This dependency on the amplitude of the HI signal happens because the {\tt GNILC} method uses the relative power of the HI signal compared to the observed signal in determining the dimension of the HI subspace. 

\begin{figure*}
  	\centering
    \includegraphics[width=\textwidth]{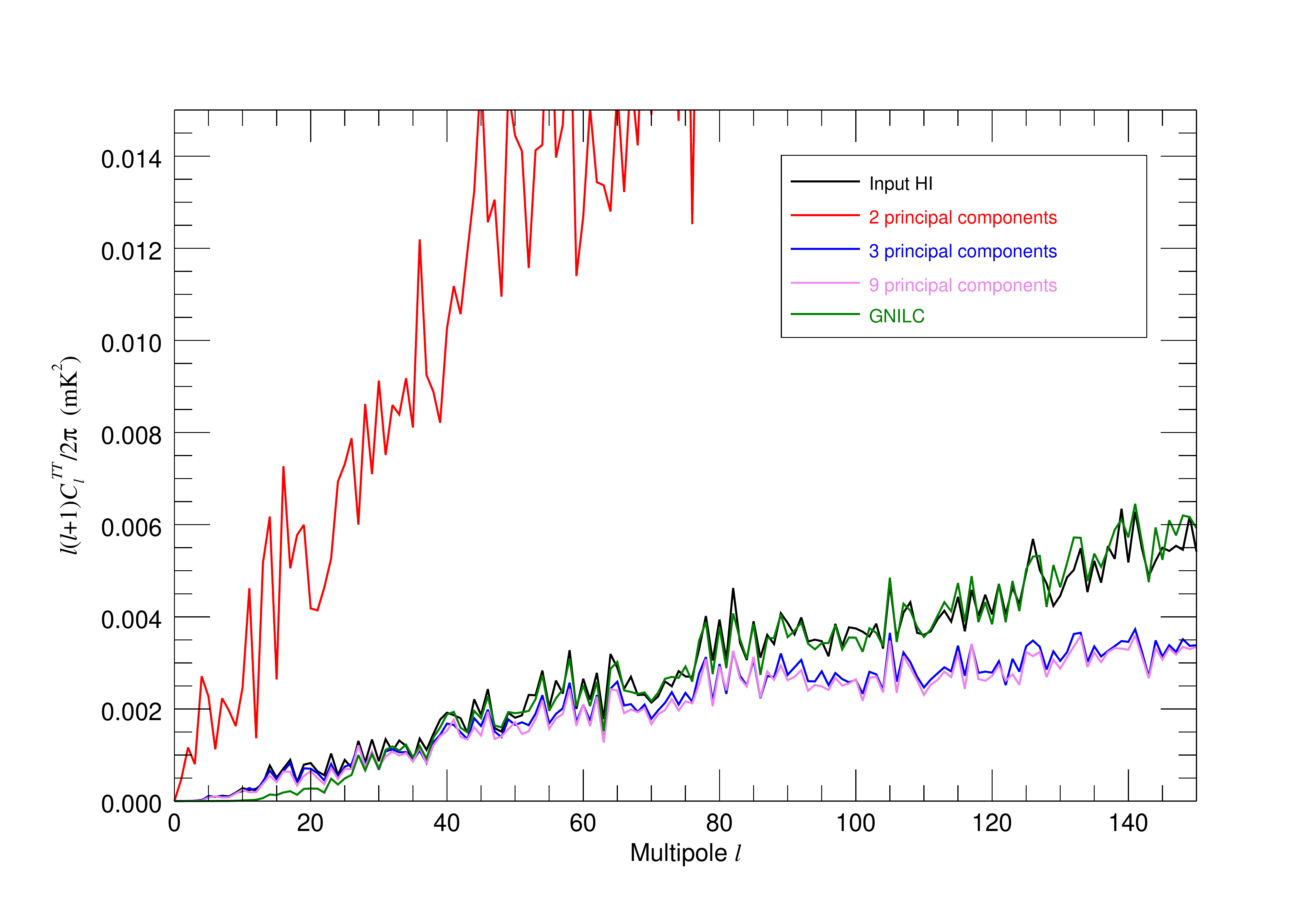}
    \caption{Comparison of {\tt GNILC} with standard PCA: power spectra, ${\ell (\ell + 1) C_{\ell} / 2 \pi}$, of the input HI signal, the {\tt GNILC} recovered HI signal, and the {\tt PCA} recovered HI signal with different number of principal components.}
    \label{fig:pca}
\end{figure*}

When our prior underestimates the HI power spectrum by a factor of 2, we cannot properly recover the HI signal. On the other hand, when the prior overestimates the HI power spectrum by a factor of 2.5, we are still able to recover the HI power spectrum with good accuracy. This difference of behavior is due to the fact that the amplitude of the HI signal is smaller than the Galactic foreground radiation by many orders of magnitude. Therefore, when the prior is overestimating the HI signal by a factor of 2.5, the amplitude of the prior power spectrum remains much smaller than the foreground radiation. Consequently, the constrained {\tt PCA} step in {\tt GNILC} algorithm estimates the same dimension for the HI subspace as the case of a prior with the correct amplitude. The same does not happen when the prior underestimates the HI signal. As now the amplitude of the HI prior is comparable to the noise, the constrained {\tt PCA} analysis needs to add an additional dimension to the foregrounds plus noise subspace at the cost of loosing part of the HI signal in this dimension. In this case, the {\tt GNILC} method cannot recover the HI signal with acceptable accuracy.

There is, however, a limit to the overestimation of the HI signal that we can adopt in the prior. The reason for this is that, in our representation of the observed data, Eq. $\eqref{diag}$, the HI subspace is characterized by the eigenvalues that are close to 1. A significant overestimation of the prior on the HI power spectrum makes the eigenvalues of the observation covariance matrix that are due to foregrounds plus noise to be of the same order of magnitude than those that are due to HI signal. In this case, the reconstructed HI power spectrum suffers from strong foregrounds contamination because the {\tt GNILC} method is not able to properly separate the HI subspace from the foregrounds plus noise subspace.

By studying different smooth priors for the HI power spectrum, we can define the range of uncertainty on the prior of the HI power spectrum that is acceptable to accurately recover the HI emission with {\tt GNILC}. We find that we can overestimate the HI power spectrum by a factor of 25 or underestimate it by a factor of 1.5 and the {\tt GNILC} method is still able to reconstruct the HI power spectrum with good accuracy (average absolute difference between the input and the reconstructed HI power spectrum normalized by the input HI power spectrum smaller than 10$\%$). 

The {\tt GNILC} method can in principle be applied to a CO intensity mapping experiment as long as the amplitude of the cosmological CO power spectrum can be known with an uncertainty of no more than 50$\%$. The current constraints on the amplitude of the CO signal are still not accurate enough to provide a reliable prior for the CO power spectrum that is needed in the {\tt GNILC} method; \citet{Li2015} compared the CO power spectra of different models available in the literature and showed that they can vary by 2 orders of magnitude.

\subsection{Comparison of GNILC with standard PCA}

We now compare the performance of the {\tt GNILC} method with the standard {\tt PCA} method. The standard {\tt PCA} method, together with the use of a transfer function to correct for HI signal loss, has been used for real data \citep{Masui2013, Switzer2013} and is commonly used in HI intensity mapping simulations \citep{Alonso2015, Bigot-Sazy2015}. For this comparison, we use simulation 3, which consists of synchrotron radiation with spatially variable spectral index, point sources, free-free radiation, and thermal noise.

\begin{table}
  \footnotesize
   \caption{The average absolute difference between the input and the reconstructed HI power spectrum normalized by the input HI power spectrum for the {\tt GNILC} method and the {\tt PCA} with 3 principal components.}
   \label{tab:pca_comp}
   \begin{center}
     \begin{tabular}{| c | c | c |}
       \hline
       Range of multipoles & {\tt GNILC} & {\tt PCA} \\ \hline
       1 - 15 & 0.93 & 0.19  \\ \hline
       15 - 30 & 0.50 & 0.12 \\ \hline
       30 - 45 & 0.09 & 0.14 \\ \hline
       45 - 60 & 0.05 & 0.15 \\ \hline
       60 - 75 & 0.06 & 0.17 \\ \hline
       75 - 90 & 0.05 & 0.22 \\ \hline
       90 - 105 & 0.04 & 0.25 \\ \hline
       105 - 120 & 0.04 & 0.27 \\ \hline
       120 - 135 & 0.06 & 0.32 \\ \hline
       135 - 150 & 0.07 & 0.38 \\ \hline
     \end{tabular}
   \end{center} 
\end{table}

In the standard {\tt PCA}, the principal components (foregrounds) are determined by looking at the largest eigenvalues of the observation covariance matrix and isolating the corresponding eigenvectors. With no prior assumption on the HI power spectrum, the optimal number of principal components detected by {\tt PCA} analysis is generally given by a constant on the whole range of angular scales considered. The {\tt PCA} thus implicitly assumes that the ratio between the HI power spectrum and the foreground power spectrum is uniform over all the angular scales, which is not the case. Therefore fixing the number of principal components as a constant on all the scales is equivalent to making a wrong assumption on the shape of the HI power spectrum. 

In our simulation 3, the standard {\tt PCA} recovers the HI signal with smallest bias when we choose the number of principal components to be equal to 3 (Fig.~\ref{fig:pca}). We note that the number of detected principal components depends on the complexity of the foregrounds and noise being simulated.

The main advantage of the {\tt GNILC} method is that, unlike the {\tt PCA}, the number of principal components is estimated locally and therefore varies with angular scale and location on the sky. The reconstructed HI power spectra for the {\tt PCA}, with different (2, 3 and 9) principal components, and for the {\tt GNILC} method are shown in Fig.~\ref{fig:pca}. In Table~\ref{tab:pca_comp}, we compare, for different ranges of multipoles, the performance of the {\tt PCA} with 3 principal components and of the {\tt GNILC} method. For each range of multipoles, we calculate the average absolute difference between the input and the reconstructed HI power spectrum normalized by the input HI power spectrum. We see that the {\tt PCA} performance is worse than the {\tt GNILC} performance by a factor of 5 on most angular scales ($\ell > 30$). For very large angular scales, $\ell < 30$, {\tt PCA} looks to better match the input HI power spectrum than {\tt GNILC}. This, however, only happens because the {\tt PCA} gives an arbitrary number of dimensions to the HI subspace on these scales that is sufficient to reconstruct the HI power spectrum with good accuracy. Note also that, as a consequence of its strictly blind analysis, the {\tt PCA} cannot have a strong confidence in the reconstructed power on these scales.

Depending on the number of principal components that are removed, the {\tt PCA} either underestimate the HI power spectrum or is strongly contaminated by residual foregrounds. It is never able to accurately measure the HI power spectrum over the whole range of scales. This is particularly true in our simulation where a large area of the sky is observed. For being large, our observed sky includes significant variations of the foregrounds with respect to the HI signal over the sky and over angular scales. The reason for the {\tt GNILC} method to reconstruct more accurately the HI power spectrum than the {\tt PCA} method for most of the angular scales is exactly what differentiates them: the number of principal components is locally determined, driven by the local signal to noise ratio, in harmonic space by the {\tt GNILC} method, while this number is fixed in all angular scales in the {\tt PCA} method. Therefore, the {\tt GNILC} method is able to consider the variations with angular scale of the foregrounds plus noise signal in reconstructing the HI signal, while the {\tt PCA} method is not.

\section{Conclusions}
\label{sec:conclusions}

In this work, we have introduced a new component separation technique for an HI intensity mapping experiment: the Generalized Needlet Internal Linear Combination ({\tt GNILC}) method. As the {\tt GNILC} method works in a wavelet space, it uses angular and spatial information to recover the HI signal from the observed data. Also, as the {\tt GNILC} method is able to explore the HI signal subspace of the observation covariance matrix, it allows us to recover the HI signal (not HI signal plus thermal noise) from the observed data.

To test the {\tt GNILC} method in a diverse set of experimental conditions, we performed several simulations for a general HI intensity mapping experiment in low redshifts. For the astrophysical foregrounds, we simulated synchrotron radiation, extragalactic point sources and free-free radiation. For the instrumental noise, we considered only thermal noise. We also studied the possibility that the experimental setup creates systematic errors in the relation between the synchrotron brightness temperature and frequency, making this relation to non-longer be a smooth function.

For the set of simulations of an HI intensity mapping experiment in low redshifts that we have considered, we were able to show that the {\tt GNILC} method is robust to the complexity of the foregrounds. We have found that we can recover the cosmological HI power spectrum for multipoles $\ell > 30$ with very good accuracy. As the number of frequency channels of the experiment increases, the {\tt GNILC} method improves in its ability to separate the HI signal from the astrophysical foregrounds and the instrumental noise. This is due to the fact that the {\tt GNILC} method is able to adapt to the number of degrees of freedom of the foregrounds plus noise signal: when the number of frequency channels increases, the {\tt GNILC} method is able to use, if necessary, more degrees of freedom to describe the foregrounds plus noise signal, resulting in a better reconstruction of the HI signal. 

To estimate locally on the sky and on the angular scales the dimensions of the foregrounds plus noise and HI subspaces, the {\tt GNILC} method uses a prior for the HI power spectrum. We have considered different incorrect priors for the HI power spectrum and studied their effect on the ability of {\tt GNILC} to recover the HI signal. Our results show that, even if we use a prior for the HI angular power spectrum to estimate the HI covariance matrix that does not have any fluctuation, the {\tt GNILC} method is still able to reconstruct the local features of the HI power spectrum. The {\tt GNILC} method, however, depends on the amplitude of the HI power spectrum. This is due to the fact that the {\tt GNILC} method uses the relative power of the HI signal compared to the observations to determine the dimension of the HI signal subspace. 


We compared the {\tt GNILC} method with the {\tt PCA} method. For all multipoles greater than 30, we recovered a more accurate HI power spectrum with {\tt GNILC} than with {\tt PCA}.

Our results show that the {\tt GNILC} method is robust to the complexity of the different astrophysical foregrounds. The {\tt GNILC} method is, therefore, a promising non-parametric component separation method for HI intensity mapping experiments. Though we have considered low redshifts for our simulations, this method can in principle be applied to high redshifts and therefore be a useful foreground cleaning method for experiments probing the Epoch of Reionization.

\section*{Acknowledgments}
LCO acknowledges funding from CNPq, Conselho Nacional de Desenvolvimento Cient\'{i}fico e Tecnol\'{o}gico - Brazil. The research leading to these results has received funding from the European Research Council under the European Union's Seventh Framework Programme (FP7/2007--2013) / ERC grant agreement no.~307209. CD also acknowledges support from an STFC Consolidated Grant (no.~ST/L000768/1). The authors also thank the anonymous referee for his/her comments on the manuscript.

\bibliography{lucas_refs}
\bsp

\appendix

\section{Needlets}
\label{app:needlets}

Wavelets are a collection of functions with properties close to those of a basis. The main difference between wavelets and a vector basis is that wavelets are redundant \citep{Daubechies1992}. Needlets were introduced as a particular construction of a wavelet family on the sphere by \citet{Narcowich2006} and \citet{Guilloux2007}. Needlets can be interpreted as a set of band-pass filters in harmonic space. The most important property of the needlets is that they have excellent localisation in the spherical harmonic domain.

We define the set of needlets windows such that, over the useful range of multipole $\ell$, we have

\begin{equation}
\label{constraint}
\sum_j [h_{\ell}^{(j)}]^2 = 1,
\end{equation} where $j$ corresponds to a specific range of multipoles and characterizes a particular needlet window. Maps of needlet coefficients are obtained, for each observed map $x(p)$, by inverse spherical harmonic transform (SHT) of the harmonic coefficients $x_{\ell m}$ of the observed map filtered by the needlet passband $h_{\ell}^{(j)}$,

\begin{equation}
\chi^{(j)}(p) = \sum_{\ell}\sum_{m=-\ell}^{\ell} x_{\ell m} h_{\ell}^{(j)} Y_{\ell m}(p),
\end{equation} where $Y_{\ell m}$ are the spherical harmonics.

For each range of multipoles $j$ and for each pixel $p$ of the corresponding needlet coefficient maps $\chi_a^{(j)}$ (one map for each frequency $a$), we can compute the covariance matrix $\mathbfss{R}$ from an average of the product of needlet coefficients. This average is done in a domain $\mathcal{D}_p$ centred at pixel $p$ and including some neighbouring pixels. This local average in pixel space, together with the property of the needlets of being almost local in spherical harmonic space, allows us to be approximately local in space and angular scale. The $ab$ component of the covariance matrix is then

\begin{equation}
\label{covwav}
\mathbfss{R}_{ab}^{(j)}(p) = \frac{1}{N_p^{(j)}} \sum_{p' \in \mathcal{D}_p} \chi^{(j)}_a(p') \chi^{(j)}_b(p').
\end{equation} The domain $\mathcal{D}_p$ is defined by convolving the product of maps with a symmetric Gaussian window in pixel space. 

For each needlet scale $j$, the number of pixels $N_p^{(j)}$, which determines the domain $\mathcal{D}_p$ used in Eq. $\eqref{covwav}$, can be constrained in the following way. When performing component separation, we may encounter the problem of ILC bias discussed in \citet{Delabrouille2009}. On small domains of the sky the statistics are computed on a reduced number $N_p$ of modes so that it can creates artificial anti-correlations between the component of interest and the contaminants. This unphysical correlations may induce a power loss in the HI signal reconstruction. The loss of the HI power is quantified by the multiplicative ILC bias $b$, derived in \citet{Delabrouille2009},
\begin{equation}
b = -\frac{n_{\mathrm{ch}} -1}{N_{m}}, 
\end{equation}
where $n_{\mathrm{ch}}$ is the number of frequency channels and $N_{m}$ is the number of modes in the domain considered. Therefore, to control the ILC bias, there is a minimum size for the set of data points on which the ILC should be implemented. We can thus constrain the number of pixels $N_p^{(j)}$ in the domain $\mathcal{D}_p$ used to compute the local covariance in such a way to maintain the ILC bias $b$ fixed for each needlet scale $j$.  



\section{Derivation of the Akaike Information Criterion}
\label{app:aic}

For a given dimension $m$ of the foregrounds plus noise subspace, we assume that the data $\textbf{\textit{x}}$ are described by a Gaussian distribution with covariance matrix $\mathbfss{R}(m) = \mathbfss{R}_{\mathrm{HI}}(m) + \mathbfss{R}_n(m)$. Therefore the likelihood distribution of the data is 

\begin{equation}
\mathcal{L}({\textbf{\textit{x}}} \vert \mathbfss{R}(m)) = \prod_{k = 1}^n \frac{1}{\sqrt{2 \pi \, \mathrm{det} \, \mathbfss{R}(m)}} \exp \left[ - \frac{1}{2} \textbf{\textit{x}}_k^T \mathbfss{R}^{-1}(m) \textbf{\textit{x}}_k \right],  
\end{equation} where $n$ is the number of modes in the needlet domain considered. The log-likelihood can be rewritten as

\begin{align}
-2 \log \mathcal{L} &= \sum_{k = 1}^n \textbf{\textit{x}}_k^T \mathbfss{R}^{-1}(m) \, \textbf{\textit{x}}_k - \log \det \mathbfss{R}^{-1}(m) + \mathrm{constant} \nonumber \\ &= n \, \mathrm{K} \left( \widehat{\mathbfss{R}}, \mathbfss{R}(m) \right) + \mathrm{constant}, 
\end{align} where $\mathrm{K} \left( \widehat{\mathbfss{R}}, \mathbfss{R}(m) \right)$ is the Kullback-Leibler divergence \citep{Kullback1951}, which measures the mismatch between the model covariance matrix $\mathbfss{R}(m)$ and the data covariance matrix $\widehat{\mathbfss{R}}$,

\begin{equation}
\label{kl}
\mathrm{K} \left( \widehat{\mathbfss{R}}, \mathbfss{R}(m) \right) = \mathrm{Tr} \left( \widehat{\mathbfss{R}} \mathbfss{R}^{-1}(m) \right) - \log \det \left( \widehat{\mathbfss{R}} \mathbfss{R}^{-1}(m) \right) - n_{\mathrm{ch}}.
\end{equation} 

For a given dimension $m$ of the foregrounds plus noise subspace, the covariance matrix ${\mathbfss{R}}(m)$ that minimizes the Kullback-Leibler divergence (or maximize the likelihood), Eq. $\eqref{kl}$, is given by the sum of Eq. $\eqref{estimate-approx}$ and Eq. $\eqref{estimate1}$.

For each location on the sky and angular scale considered, we can then select the best rank value $m_b$ for the foregrounds plus noise covariance matrix by minimizing the Akaike Information Criterion (AIC),

\begin{equation}
\label{aic}
\mathrm{AIC}(m) = 2 \, n m  - 2 \log (\mathcal{L}_{\mathrm{max}}(m)).
\end{equation} Since the transformed data covariance matrix can be diagonalized as $\widehat{\mathbfss{R}}_{\mathrm{HI}}^{-1/2} \widehat{\mathbfss{R}} \widehat{\mathbfss{R}}_{\mathrm{HI}}^{-1/2} = \mathbfss{U} \mathbfss{D} \mathbfss{U}^T$, where, as in Eq. $\eqref{diag}$,

\begin{equation}
\mathbfss{U} =[\mathbfss{U}_N \; \mathbfss{U}_S ] \;\;\; \mathrm{and} \;\;\; \mathbfss{D} = \begin{bmatrix} 
\widehat{\mathbfss{D}}_N & 0 \\
0 & \widehat{\mathbfss{D}}_S
\end{bmatrix}  
\end{equation} and the maximum likelihood happens at $\widehat{\mathbfss{R}}_{\mathrm{HI}}^{-1/2} \mathbfss{R} \widehat{\mathbfss{R}}_{\mathrm{HI}}^{-1/2} = \mathbfss{U}_N (\mathbfss{D}_N - \mathbfss{I}) \mathbfss{U}_N^T + \mathbfss{I}$, we have that

\begin{align}
\label{log}
- 2 \log (&\mathcal{L}_{\mathrm{max}}(m)) =  \nonumber \\ &= n \, \mathrm{K} \left( \widehat{\mathbfss{R}}_{\mathrm{HI}}^{-1/2} \widehat{\mathbfss{R}} \widehat{\mathbfss{R}}_{\mathrm{HI}}^{-1/2}, \mathbfss{U}_N (\mathbfss{D}_N - \mathbfss{I}) \mathbfss{U}_N^T + \mathbfss{I} \right) \nonumber \\ &= n \, \mathrm{K} \left( \mathbfss{U} \mathbfss{D} \mathbfss{U}^T, \mathbfss{U}_N (\mathbfss{D}_N - \mathbfss{I}) \mathbfss{U}_N^T + \mathbfss{I} \right) \nonumber \\ &= n \, \mathrm{K} \left( \mathbfss{D}, \mathbfss{U}^T \left[ \mathbfss{U}_N (\mathbfss{D}_N - \mathbfss{I}) \mathbfss{U}_N^T + \mathbfss{I} \right] \mathbfss{U} \right) \nonumber \\  &= n \, \mathrm{K} \left( \begin{bmatrix} 
\widehat{\mathbfss{D}}_N & 0 \\
0 & \widehat{\mathbfss{D}}_S
\end{bmatrix},
\begin{bmatrix} 
\widehat{\mathbfss{D}}_N & 0 \\
0 & \mathbfss{I}
\end{bmatrix} 
\right) \nonumber \\ &= n  \left( \mathrm{Tr} \left( \begin{bmatrix} 
\mathbfss{I} & 0 \\
0 & \widehat{\mathbfss{D}}_S
\end{bmatrix}
\right)
- \log \det \left( \begin{bmatrix} 
\mathbfss{I} & 0 \\
0 & \widehat{\mathbfss{D}}_S
\end{bmatrix}
\right) - n_{\mathrm{ch}} \right) \nonumber \\
&= n \left( m + \sum_{i = m + 1}^{n_{\mathrm{ch}}} \mu_i - \sum_{i = m +1}^{n_{\mathrm{ch}}} \log \mu_i - n_{\mathrm{ch}} \right) \nonumber \\ &= n \sum_{i = m + 1}^{n_{\mathrm{ch}}} [\mu_i - \log \mu_i - 1],
\end{align} where $\mu_i$ are the eigenvalues of the transformed covariance matrix of the observed data. Substituting Eq. $\eqref{log}$ in the expression for the AIC, Eq. $\eqref{aic}$, we obtain 

\begin{equation}
\label{aic1}
\mathrm{AIC}(m) = n \left(2m + \sum_{i=m+1}^{n_{\mathrm{ch}}} [\mu_i - \log \mu_i - 1] \right).
\end{equation} The dimension of the foregrounds plus noise subspace $m$ is then estimated by minimizing Eq. $\eqref{aic1}$ in each region considered. 

\label{lastpage}

\end{document}